\documentclass{article}
\usepackage{arxiv}

\usepackage[utf8]{inputenc} 
\usepackage[T1]{fontenc}    
\usepackage{hyperref}       
\usepackage{url}            
\usepackage{booktabs}       
\usepackage{amsfonts}       
\usepackage{nicefrac}       
\usepackage{microtype}      
\usepackage{lipsum}		
\usepackage{graphicx}
\usepackage{natbib}
\usepackage{doi}
\usepackage{amsmath}

\usepackage{amssymb,amsfonts,amsthm,bm,bbm}
\usepackage{algorithm}
\usepackage{soul}
\usepackage{algorithmic}
\usepackage{array}
\usepackage{graphicx}
\usepackage{booktabs}
\usepackage{multirow}
\usepackage{pifont}
\usepackage{tcolorbox}
\usepackage{enumitem}
\usepackage{multicol}

\renewcommand{\paragraph}[1]{\smallskip\noindent\textbf{#1}}
\newcommand{\prpd}{\textsc{pRPD}}
\newcommand{\hp}{\textsf{HP}}
\newcommand{\rp}{\textsf{RP}}
\newcommand{\pd}{\textsf{Protocol Descriptor}}
\newcommand{\ap}{\textsf{AP}}
\newcommand{\eo}{\textsf{EO}}
\newcommand{\daa}{\textsc{Difficulty Altering}}
\newcommand{\smb}{\textsc{Selfish Mining with Bribing}}
\newcommand{\pcf}{\textsc{Quick Fork}}
\newcommand{\tw}{\textsc{Transaction Withholding}}
\newcommand{\gossip}{$\Pi_{gossip}$}
\newcommand{\txincl}{$\Pi_{tx-inclusion}$}
\newcommand{\proName}{\textcolor{olive}{\textbf{PRAGTHOS}}}

\newcommand{\myitem}[1]{%
  \item\makebox[1.5em][l]{#1}%
  }

\newtheorem{theorem}{Theorem}[section]
\newtheorem{corollary}{Corollary}[section]
\newtheorem{claim}{Claim}[section]
\newtheorem{lemma}[theorem]{Lemma}
\theoremstyle{definition}
\newtheorem{definition}{Definition}[section]
\setlist[itemize]{leftmargin=3mm}

\title{PRAGTHOS: PRActical Game THeOretically Secure Proof-of-Work Blockchain}

\date{February 13, 2023}	

\author{ \href{https://orcid.org/0000-0002-5662-0386}{\includegraphics[scale=0.06]{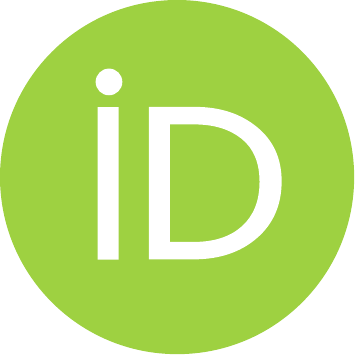}\hspace{1mm}Varul Srivastava} \\
	Machine Learning Lab\\
	IIIT Hyderabad\\
	\texttt{varul.srivastava@research.iiit.ac.in} \\
	\And
	\href{https://orcid.org/0000-0003-4634-7862}{\includegraphics[scale=0.06]{orcid.pdf}\hspace{1mm}Dr. Sujit Gujar} \\
	Machine Learning Lab\\
	IIIT Hyderabad\\
	\texttt{sujit.gujar@iiit.ac.in} \\
}

\renewcommand{\shorttitle}{PRAGTHOS}

\hypersetup{
pdftitle={PRAGTHOS},
pdfsubject={q-bio.NC, q-bio.QM},
pdfauthor={David S.~Hippocampus, Elias D.~Striatum},
pdfkeywords={First keyword, Second keyword, More},
}

\begin{document}
\maketitle

\begin{abstract}
Security analysis of blockchain technology is an active domain of research. There has been both cryptographic and game-theoretic security analysis of  Proof-of-Work (PoW) blockchains. Prominent work includes the cryptographic security analysis under the Universal Composable framework and Game-theoretic security analysis using Rational Protocol Design. These security analysis models rely on stricter assumptions that might not hold.

In this paper, we analyze the security of PoW blockchain protocols. We first show how assumptions made by previous models need not be valid in reality which an attacker can exploit to launch attacks that these models fail to capture. These include (1) Difficulty Alternating Attack, under which forking is possible for an adversary with $< \frac{1}{2}$ mining power, (2)  Quick-Fork Attack, (3) a general bound on selfish mining attack and, (4) transaction withholding attack. Following this, we argue why previous models for security analysis fail to capture these attacks and propose a more practical framework for security analysis -- \prpd. We then propose a framework to build PoW blockchains \proName, which is secure from the attacks mentioned above. Finally, we argue that PoW blockchains complying with the \proName\ framework are secure against a computationally bounded adversary under certain conditions on the reward scheme.

\end{abstract}

\section{Introduction}
\label{sec:intro}



Blockchain technology was introduced by Satoshi Nakamoto~\cite{nakamoto-2008} via Bitcoin -- an alternative to centralized financial institutions. Blockchain is an immutable \emph{distributed} ledger that achieves distributed trust. It stores the data in blocks connected through certain cryptographic links and ensures consensus on a single state (ordering of blocks) across all participants. Vitalik~\cite{Ethereum-WP} introduced the Ethereum blockchain, which supports Turing complete smart-contract functionality~\cite{Szabo-1997-SmartContract}. Since then, blockchain technology has found immense applications in multiple domains, such as financial, supply-chains, Smart-city and IoT systems, Decentralized Governance, Voting and Auction systems, etc. Blockchain technology \emph{distributes} trust by maintaining \emph{consensus} over the state of the system.

The FLP impossibility theorem (\cite{FLP-1983}) states that a deterministic algorithm cannot achieve consensus in an asynchronous system even if a single node is faulty. Blockchains are \emph{decentralized} systems that use \emph{incentive-engineering} to overcome these impossibilities. It uses \emph{Proof-of-Work} (PoW) consensus algorithm to ensure consistency and overcome such impossibilities otherwise existent in distributed systems. In PoW, each participating node (miner) is supposed to solve a certain cryptographic puzzle through queries to write to the blockchain. 
The query is successful if the block's hash value is lesser than some decided target. Blockchain protocols overcome impossibilities because (i) they are incentive-based protocol -- offers rewards to the miner who writes the block (ii) they are non-deterministic. 

As blockchain protocols gained popularity, researchers have done extensive security analysis of PoW-based blockchains~\cite{Garay-2015-Backbone, Garay-2017-chainsVariable, Eyal-2014-Selfish}. Badertscher et al.~\cite{Badertscher-2017-BitcoinUC} use the \emph{Universal Composable (UC) Framework} (introduced by Canetti~\cite{Canetti-2001-UC}) to propose a universal composable treatment of Bitcoin. However, in the security analysis of blockchains, we should consider not only \emph{cryptographic} but also \emph{game-theoretic} security. Garay et al.~\cite{Garay-2013-RPD} proposed \emph{Rational Protocol Design} (RPD) for game-theoretic security analysis of any distributed cryptographic protocol. Badertscher et al.~\cite{Badertscher-2018-BitcoinRPD,Badertscher-2021-Bitcoin51} modified the RPD model for game-theoretic security analysis of PoW blockchains. There has been extensive research in the domain of game-theoretic security of PoW blockchain~\cite{Judmayer-2021-AIM, Katz-2017-Whale1, Katz-2017-Whale2, b15, b16,shoeb,chen,b14,b21}. In particular, Badertscher et al.~\cite{Badertscher-2018-BitcoinRPD} proves that Bitcoin satisfies strong notions of security such as\emph{strong-attack-payoff-security}\footnote{see Definition~\ref{def:strong-aps} in this paper} proposed in~\cite{Garay-2013-RPD} under honest majority assumption. However, this work demonstrates how an attacker can attack Bitcoin or many PoW blockchains even if it has $<50\%$ computing power using \daa\ attack. 

The primary reason for the existing frameworks could conclude security is that they assume one or more of the following: (i) constant conversion rate of crypto-currency to fiat currency, (ii) constant block reward, or (iii) constant difficulty. In reality, these are dynamic parameters.
Additionally, these works categorize miners into \emph{honest party} (\hp) -- who follow the protocol honestly or \emph{advesarial party} (\ap)-- who launch attacks on the protocol. Some miners typically follow the protocol but may deviate conditionally if a strategy offers a higher utility without disrupting the blockchain. We call such miners \emph{Rational Party} (\rp). In this work, we show that multiple attacks are possible in the presence of \rp\ on the PoW blockchain, which RPD-based models fail to capture. Hence, our goal is to build a practical model that aligns with the real-world environment for game-theoretic security analysis of a PoW blockchain exists. 
\subsection*{Our Contributions}
\label{ssec:our-contributions}
Our contributions in this work are threefold. (i) We identify specific attacks on existing PoW blockchain protocols and explain why existing security models fail to capture them. (ii) We propose a new model -- \textsc{Practical Rational Protocol Design} (\prpd) to overcome the limitations of existing models for PoW blockchain. (iii) We propose solutions to overcome the attacks identified, propose a novel framework \proName\, and perform security analysis of PoW blockchains complying with the framework. 

\noindent \textbf{Attacks:} For PoW blockchains, we show that there exist previously undiscovered attacks. Additionally, on known attacks, we prove that the fraction of adversarial computing power required to launch the existing attack is much smaller than the existing bounds.
\begin{itemize}
    \item We prove that there exist \daa\ in which \ap\ with $< \frac{1}{2}$ can launch a successful fork without the help of \rp. The previous estimate of this bound was $\frac{1}{2}$ ($51\%$ attack). As a corollary, such an attack is possible in Bitcoin even when \ap\ controls $45\%$ of the mining power (Theorem~\ref{thm:daa-attack}).
    \item We show that there exists a security attack \pcf\  where \ap\ leverages incentive-driven deviations from \rp\ (Theorem~\ref{thm:pcf-attack}). 
    \item We also show that the bound for selfish-mining proposed in~\cite{Eyal-2014-Selfish} is a special case of a more general bound. Using our proposed model \prpd, we can capture this more general bound when the selfish-mining attack is combined with whale-transactions~\cite{Katz-2017-Whale2}, which we call -- \smb.
    \item We show that under transaction-fee only model (TFOM)\footnote{TFOM formally explained in Appendix~\ref{app:misc-def}}, there exists a deviation from \emph{Gossip} protocol (Lemma~\ref{lemma:gossip-subopt}) which we call a transaction-withholding attack. This strategic deviation can either lower the throughput of the PoW blockchain or centralize the protocol.
\end{itemize}

\noindent \textbf{\prpd:} We discuss why previous works fail to capture these security attacks and make corresponding modifications to introduce a more practical model for security analysis of PoW blockchain protocol -- \prpd. In this model, we account for multiple mining rounds while also capturing the system's dynamics -- variable block reward and variable difficulty. We also capture the external responses to the system through externalities (change in conversion rate to fiat currency), which depend on the strategies followed by the miners. Further, we include three types of players, altruistic (honest), conditionally-deviating (rational), and deviating (adversarial) in our analysis.

\noindent \textbf{Detering Attacks \& \proName:} To rectify the security attacks above, we propose certain amendments that a PoW blockchain protocol should adopt. Combining these modifications, we propose a novel framework to build a PoW blockchain, namely \proName. 
\begin{itemize}
     \item We determine appropriate hyper-parameter values to ensure \daa\ attack is not possible (Corr.~\ref{corr:daa-sec}). 
    \item We propose \texttt{PC-MOD} as a deterrence to \pcf\ attack. We show that the existence of such a protocol (to be run in case \pcf\ attack is launched) is sufficient to prevent the attack. 
     \item We also show a pessimistic result that there does not exist any PoW blockchain protocol that is attack-payoff secure against Selfish mining (Theorem~\ref{thm:smb-impossibility}).
    \item We propose \txincl\ protocol to ensure that deviation from \emph{Gossip} protocol is disincentivized.
    \item Based on these modifications, we introduce \proName\ and perform security analysis.
    \begin{itemize}
        \item Under inflationary block-reward scheme we show that \proName\ is \emph{strongly attack-payoff secure} (Theorem~\ref{thm:spas}).
        \item We show that under the deflationary block-reward scheme: (1) no PoW blockchain protocol is \emph{strongly attack-payoff secure} against a general adversary (Theorem~\ref{thm:dra}). (2) \proName\ is secure against an adversary with an attack strategy bounded in the number of rounds (Theorem~\ref{thm:semi-major-theorem}).
    \end{itemize}
\end{itemize}

With \prpd\ and \proName, we have significant results : (i) Bitcoin is not secure even with the honest majority, (ii) Crypto-currencies with deflationary reward schemes are not strongly attack-payoff secure. Thus, we believe this work lays the foundations for building future PoW blockchain protocols. 

\subsection*{Related Work}
\label{ssec:related-work}

The analysis of cryptographic and game-theoretic security of 
PoW blockchain protocols have been extensively done in recent literature, and multiple attacks have been discovered and discussed. In this section, we contrast our work with the existing attacks and the analysis frameworks. 
%
%

\paragraph{\textbf{General Security.}} Tsabary and Eyal discuss in~\cite{b16} that if the subsidy (cumulation of transaction fee and block-reward) is small enough and the cost of mining is high enough, then the gap between intervals where miners mine blocks increases. Further, this gap varies based on the relative size of mining pools, and it is incentivized to form coalitions to increase the reward. This means that sparse transaction distribution and lesser block rewards lead to the centralization of the system.  
Wei et al.~\cite{b17} study the effect of network delays in Blockchain networks. The authors propose a more general framework to model network delays, with large possible delays happening probabilistically. However, in their analysis, the adversary is restricted to performing network delays. Gazi et al.~\cite{practicalSettlement2021} proposed an efficient method to compute bounds on settlement time for PoW blockchain given computation power and network delays. 
Yuan et al.~\cite{b20} discuss  the protocol with $(1 + \mu)$ miners, out of which $\mu$ are corrupted. Further, the adversary can delay the network by at most $\frac{1}{np\Delta}$ with probability $1$. In this case, the authors discuss conditions under which chain growth, chain quality, and common prefix properties hold\footnote{there $3$ properties are required for blockchain security}. Momeni et al.~\cite{Momeni-2022-Fairblock} propose an identity-based encryption mechanism to prevent front-running. However, this solution is not efficiently applicable in public PoW blockchains because the protocol is based on the committees` identities that perform consensus.  
 There exist works like~\cite{Badertscher-2017-BitcoinUC, GrafDSUC21} which perform universal composable treatment of Blockchains. The general framework of cryptographic security analysis discussed in~\cite{GrafDSUC21} is applicable to general distributed ledgers, whereas our work analyses game-theoretic security as well.
Ke et al.(\cite{b18}) model contingency plans for severe attacks on blockchains. They propose a framework that inputs the attack on the blockchain and gives the contingency plan, detection plan, and level of damage as output. Their work proposes frameworks for contingency plans and assessing the level of damage for the given attack, unlike our work which focuses on modeling the system to capture all possible attacks and proposing a framework to resolve those attacks.

\paragraph{\textbf{Rational Security.}} Han et al.~\cite{b15} points out that the honest majority assumption is not true, especially for some miners who may be adversaries who can be lured to work on a fork through incentive engineering. The authors show that there exist incentive structures if we consider the existence of multiple blockchains in the system, such that miners can be incentivized to launch $51\%$ attacks on the blockchain with less computing power. Their work differs from ours because they consider systems with multiple blockchains. We discuss multiple attacks, which are possible even in a blockchain system where the majority of miners are honest. Kevin and Katz (\cite{Katz-2017-Whale2},\cite{Katz-2017-Whale1}) performed one of the earliest works which discuss the idea of (non-adversarial) \rp\ who can deviate from the longest chain. Their attack is through \emph{Whale Transactions} -- transactions with abnormally high transaction fees. Judmayer et al. (\cite{Judmayer-2021-AIM,b15}), consider an adversary who bribes \rp\ through external guaranteed incentives through smart contracts for joining the attack -- the authors call it \emph{Algorithmic Incentive Manipulations}. 
Our model uses \rp\ to capture a larger pool of miners as they still may deviate even in the absence of bribes. Thus, our model is more general. Another line of attacks is when Block Reward reduces to become negligibly small and miner incentive is driven by Transaction Fees only. Most of the existing works~\cite{b14,shoeb,transactionFee-FC-2015,transactionFeeQueuing,TxnFeeEconomics,Roughgarden2021TransactionFM} either study the incentive-based deviations of transaction proposers, or makes optimistic assumptions on the distributions of Transactions. Badertscher et al.~\cite{Badertscher-2018-BitcoinRPD} and Carlsten et al.~\cite{Narayanan} show that under an unfavorable (but equally possible) distribution of incoming transactions, the Protocol is not secure. We discuss a possible deviation that compromises the protocol under such transaction distribution and how to tackle such deviation. 

\paragraph{\textbf{Security Models.}} Badertscher et al.~\cite{Badertscher-2018-BitcoinRPD, Badertscher-2021-Bitcoin51} performed RPD on Bitcoin and analysis on double-spending in case of $51\%$ attack. These works fail to capture attacks that enable Forking and Double-spending for an adversary with $\beta_{adv} < \frac{1}{2}$ without leveraging deviations from other miners. The utility model in~\cite{Badertscher-2018-BitcoinRPD} fails to capture variable block reward, variable conversion rate, and the variable cost of mining. The authors generalized the utility model in RPD  to capture variable block reward and cost of mining in~\cite{Badertscher-2021-Bitcoin51}. The authors argue that due to decreasing block-reward, \emph{attack-payoff-security} should be considered only for \emph{finite-horizon}. As the conversion rate (from cryptocurrency to fiat) is also dynamic (dependent on participating players' strategies and supply of the cryptocurrency). Note that this conversion rate can increase\footnote{as is visible for Bitcoin and Ethereum's historical prices}, rendering the finite horizon argument made in~\cite{Badertscher-2021-Bitcoin51} impractical. On the other hand, our model captures both discounting and deflationary block rewards. Thus, our results capture a more realistic behavior of the system.

\section{Preliminaries}
\label{sec:prelims}
In this section, we discuss the background, notations, and formalism necessary for the discussions that follow in this paper.
\subsection{Universal Composability}
\label{ssec:uc-prelims}
First, we brief the set of protocols that are considered when defining security for PoW blockchains. 
Canetti~\cite{Canetti-2001-UC} proposes the framework of \emph{Universal Composability} (UC), and later \cite{Badertscher-2017-BitcoinUC} proposes a UC treatment of PoW blockchain. Similar to the previous works, our model assumes all participating parties are \emph{Probabilistic Polynomial Time} (PPT) \emph{Interactive Turing Machines} (ITMs). 
 
\subsubsection{$\mathcal{G}_{ledger}$ Ideal Functionality} 
\emph{Ideal Functionality} is the functionality that the proposed protocol aims to achieve. 
A PoW blockchain aims to achieve $\mathcal{G}_{ledger}$ functionality as described in~\cite{Badertscher-2017-BitcoinUC}. This functionality stores a ledger that maintains the system's state, with each state transition performed through transactions submitted by participating parties. Different parties might have different points of view for the state head; therefore, the functionality stores a pointer for the state head according to each party. The \texttt{VALID-TX} predicate checks the validity of the transactions and appends them to the ledger if valid (makes corresponding state transactions). The predicate enforces \emph{ledger growth}, \emph{chain quality}, and \emph{transaction liveness}.

\subsubsection{$\mathcal{G}_{weak-ledger}$ Relaxed Functionality}
The concept of relaxing the ideal functionality exists because there are multiple attacks and other behaviors in the real-world which need to be simulated in the ideal world. The relaxed, ideal world functionality of $\mathcal{G}_{ledger}$  is defined as, $\mathcal{G}_{weak-ledger}$ in~\cite{Badertscher-2018-BitcoinRPD}. The relaxed functionality stores the states as a tree instead of a linked list, which allows forks of arbitrary lengths to exist, and we choose the longest chain out of these forks, as in all bitcoin-like PoW blockchains. 

Further, since in real world it is indistinguishable if the block was mined by an adversarial, rational or honest miner, the functionality relaxes checks on the chain quality and chain growth properties, which are verified in
\texttt{StateExtend} policy. In $\mathcal{G}_{weak-ledger}$ the \texttt{StateExtend} policy is relaxed to \texttt{WeakStateExtend}. The relaxed functionality accepts all transactions in the state buffer without checking their validity against the state-tree because these transactions might be valid in one of the multiple chains of the state-tree. In addition, the ability to create forks can be invoked by the \texttt{Fork} command, which extends the chain from an indicated block, instead of the traditional \texttt{Next Block} command, which extends the chain from the header for the calling party. (For more details on $\mathcal{G}_{ledger}$  and $\mathcal{G}_{weak-ledger}$, please refer to~\cite{Badertscher-2017-BitcoinUC} and~\cite{Badertscher-2018-BitcoinRPD} respectively.)

\subsection{PoW Blockchain Protocol}
\label{ssec:pow-protocol}
A PoW blockchain consists of a chain and a hash-pointer-based linked list. Further, to add a block, we have to include the hash of the parent block to which it points and solve a PoW puzzle (by choosing a nonce) such that the block's hash is less than a target. A lower target indicates the greater difficulty of the puzzle. We represent a PoW blockchain that UC-realizes (refer Def. 20,~\cite{Canetti-2001-UC}) the relaxed ideal functionality as $\Pi$.

Let $C_{i}^{\lfloor k}$ be the subset of all but the last $k$ blocks in chain $C_{i}$. We consider additional and practical relaxation of $G_{ledger}$ ideal functionality, namely variable difficulty. The probability of mining a block on different chains can vary based on the chain's difficulty. The difficulty of mining a block on any $C_{i}$ is publicly known, as it can be calculated using the publically available ledger. The sub-protocol $isValidStruct(C)$, defined in Section 5.1 of~\cite{Badertscher-2017-BitcoinUC}, takes the difficulty parameter corresponding to the chain given as input instead of the system parameter. The system rewards miners with a block reward for maintaining a ledger by solving PoW puzzles.
\subsubsection{Ledger}
\label{sssec:ledger}
The ledger contains blocks that a miner mines. The block $b = (h_{parent}, nonce,$ $txMD,\{0,1\}^{*})$ contains (i) hash pointer to the parent block $h_{parent}$, (ii) a random nonce $nonce$ and, (iii) the meta-data for the set of transactions to be included. This metadata depends upon the blockchain implementation details. E.g., it includes the root of the Merkle tree containing a set of transactions, which in turn also include the coinbase transaction\footnote{Bitcoin's mechanism to pay miners a block reward} in Bitcoin~\cite{nakamoto-2008}. Further, the contract can include other information in the form of arbitrary binary strings to enable additional features in the ledger. 

\subsubsection{Timesteps}
\label{sssec:timesteps}
Since there is no global clock in the system (which runs in a partially-synchronous setting~\cite{partSync-1984}), we cannot divide the measurement of events in terms of time. We use the notion of \emph{rounds}, which depends on the number of hashes computed by the system. 
\begin{definition}[Round]\label{def:round}
    A \emph{round} is a duration in which any miner makes $q$ queries\footnote{in PoW blockchain these are hash-queries}.
\end{definition}
In addition to \emph{rounds}, we define an \emph{epoch} to capture the event of the change in difficulty of mining and \emph{phase} to denote the change in block-reward (e.g. in Bitcoin, reward halves every $210,000$ blocks).
\begin{definition}[Epoch]\label{def:epoch}
    An \emph{epoch} is a duration in which the system mines $\lambda$ blocks at constant difficulty. The difficulty is scalled by $\tau$ at the end of an epoch for $\tau\in [\tau_{min}, \tau_{max}]$.
\end{definition}
\begin{definition}[Phase]\label{def:phase}
    A \emph{phase} is a duration in which the system mines $\Lambda$ blocks. At the end of a phase $j$, block reward gets updated by the update rule $r_{block}^{new} \gets \varrho(r_{block}^{old},j)$\footnote{for Bitcoin $\varrho(r_{block}^{old},j) = \frac{r_{block}^{old}}{2}$} for some $\varrho:\mathbb{R}\times\mathbb{N}\rightarrow\mathbb{R}$
\end{definition}
The protocol execution comprises different parties participating in each round, whose roles are explained below.
\subsubsection{Parties}
\label{sssec:parties}
PoW-based blockchains have three types of parties -- Altruistic (Honest), Adversarial and Rational. Each party consists of a set of miners who exhibit the same behavior. In our formulation, to model market responses, we introduce a dummy, passive party, an \emph{External observer}.\\

\noindent \textbf{External Observer(\eo)} acts as an observer of public chains of the blockchain. We assume a \emph{tolerance factor} $\rho$ to account for network latency and accidental forks. That is, \eo\ considers a forked chain $C_{f}$ as a forking (security) attack only if there existed another chain $C_{h}$ which previously had a lead of $\rho$ blocks and $C_{f}$ overtakes $C_{h}$ eventually. In this work, we model the drastic conversion rate changes that can happen within one round and not the slow ones which we observe.  In round $t$, \eo\ sends a signal $\theta(t)$ to the parties. If $\theta(t)=1$  implies that there is no major disruption in the conversion rate. 
\begin{equation}\label{eqn:externality}
\theta(t) = 
    \begin{cases}
        1 & \text{if all follow protocol honestly}\\
        e_{fairness} & \text{if fairness attacks}\\
        e_{security} & \text{if security attacks}
    \end{cases}
\end{equation}

\begin{itemize}
        \item[1] \textbf{Honest Party(\hp):} These miners control $\beta_{hon}$ fraction of total mining power. \hp\ participate in the system by following the PoW blockchain protocol honestly if participation is profitable; otherwise, they do not participate. 
        \item[2] \textbf{Adversarial Party(\ap):} These miners control $\beta_{adv}$ fraction of the total mining power. Adversarial miners launch attacks and deviate from the honest protocol. They received payoff through the crypto-currency as well as by short-selling\footnote{short-selling is typical terminology in stock-trading; one sells the stocks (in this case, cryptocurrency) not owned by it and repurchases at a lower price in the future.}. Thus the change in $\theta(t)$ is inconsequential for \ap.
        \item[3] \textbf{Rational Party(\rp):} These miners control $\beta_{rat}$ fraction of total mining power and follow the protocol --  $\Pi$ unless there exists a deviation with higher utility. However, this party would deviate only if the deviation cannot be a security attack observed by \eo, i.e., it is guaranteed that after deviation $\theta(t)\neq e_{security}$.
    \end{itemize}

Note that we consider all miners to be computationally bounded and have a fixed computing power. However, if a single player increases its mining power, we consider it multiple miners (without loss of generality); each making $q$ queries in a round. 

Next, we explain the role of \eo\ in modeling the market response, conversion rates, etc.

\subsubsection{Modelling Externalities}
\label{sssec:externality}
\noindent In \prpd, externality plays an important role while modeling the behavior of rational players. Badertscher et al. ~\cite{Badertscher-2018-BitcoinRPD} assign a very high (exponential in poly-log of security factor) negative payoff to the \pd\ to model the effect on the protocol (or the value of the cryptocurrency). In our case, we have defined the term $\theta(t)$, which changes the crypto-currency value based on the strategies miners follow, as observed by \eo\ -- the value reduces to $< 1$ if there is deviation. If the payoff in cryptocurrency is $\mathcal{R}$, then the actual payoff is $\theta(t)CR\mathcal{R}$. We model the market response to different strategies through $\theta(t)$ and the supply-demand fluctuation is handled by $CR$. We divide the attacks on the blockchain system into two categories.

\begin{itemize}
    \item[1] \textbf{Fairness-Attacks} compromise on fairness, i.e., some miner gaining more rewards than its fair share or some user needing to wait for a transaction to be accepted unreasonably high. In these attacks, the security remains intact. We assume if such an attack is observed by \eo, $\theta(t) = e_{fairness}<1$. 
    \item[2] \textbf{Security-Attacks} is when security of the protocol is compromised. In this case, the externality parameter $\theta(t) = e_{security}$.\footnote{$e_{fairness} >> e_{security}$ because security attacks are a more serious threat.}
\end{itemize}
 For example, security threat can be modeled as $e_{security} = negl(\kappa)$. In that case, $e_{security} = \frac{1}{poly(\kappa)}$. 

\subsection{Existing Attacks on PoW Blockchain}
\label{ssec:attacks-prelims}
There have been several attacks proposed \cite{Eyal-2014-Selfish, Katz-2017-Whale1, Katz-2017-Whale2, Eskandari-2019-frontRunning, Breidenbach-2018-HydraFR, Kalodner-2015-NamecoinFR} for PoW based blockchain. We enlist the previously discovered attacks which are relevant to our work.

\subsubsection{Selfish Mining}
\label{sssec:selfish-mining-eyal}
\emph{Selfish mining} is an attack proposed by Eyal and Sirer~\cite{Eyal-2014-Selfish}. In this attack, the adversary gains a higher fraction of the total block reward for any continuous set of blocks mined than the fraction of total mining power held by the adversary. Therefore, this is an attack on the system's fairness, not a security attack. However, the authors of~\cite{Eyal-2014-Selfish} show that after certain modifications, only an adversary with $> \frac{1}{4}$ mining power can successfully launch a selfish mining attack.
\begin{claim}[\cite{Eyal-2014-Selfish}, Observation 1]
For a given $\gamma$, an adversarial pool of size $\beta_{adv}$ obtains a revenue larger than the relative size for $\beta_{adv}$ in the following range: 
\begin{equation}\label{eqn:selfish-mining-eyal}
\frac{1-\gamma}{3 - \gamma} < \beta_{adv} < \frac{1}{2}
\end{equation}
Where $\gamma$ is the fraction of non-adversarial parties that mine on top of the adversarial block in case of competition among two blocks for the longest chain. 
\end{claim}
Carlsten et al.~\cite{Narayanan} discuss the possibility of Selfish Mining being profitable for an adversary with arbitrarily low mining power. Their result holds only in Transaction Fee Only Model (TFOM-- when block rewards are negligible) and relies on the non-uniformity of rewards. Our work proposes that in the presence of \rp, an adversary with arbitrarily low mining power might be incentivized to launch a Selfish mining attack and is possible when there are block significant rewards.
\subsubsection{Forking Attack}
\label{sssec:forking-attack}
A forking attack is a security attack on the PoW blockchain. We say a chain $C_{A}$ overtaking chain $C_{H}$ at some time is a forking attack if (1) at some time $t_{1}$, $length(C_{H} - C_{A}) \geq k$ (where $k$ is a parameter set by the PoW blockchain\footnote{for Bitcoin $k = 6$}. (2) $C_{A}$ overtakes $C_{H}$ at some time $t_{2} > t_{1}$. Such attacks can lead to double-spending and are, therefore, a serious security threat to the blockchain. 

\subsubsection{Timewarp Attack}
\label{sssec:timewarp-attack}
Several adversarial manipulations exist such as the Timewarp-attack~\cite{timewarp-bad} that uses incorrect time-stamping to reduce mining difficulty. However, Timewarp-attack is feasible only if adversary holds $> 51\%$ of the mining power. PoW blockchains with difficulty recalculation each round (like the Verge~\cite{verge}) suffer from security threat. However, PoW blockchains like the Bitcoin are secure against timewarp attack under honest majority. In contrast, the \daa\ attack (Section~\ref{ssec:bz-attacks}) are possible in Bitcoin and similar blockchains even if majority is honest. 

\subsubsection{Front-running attacks}
\label{sssec:front-running}
We assumpe a special class of adversarial PPT ITMs $\mathcal{A}_{fr}^*$ -- \emph{front-running} adversaries, originally defined in~\cite{Badertscher-2018-BitcoinRPD} 
\begin{definition}[Front-Running, Def. 2,~\cite{Badertscher-2018-BitcoinRPD}]\label{def:front-running}
An adversary $A \in \mathcal{A}^{*}_{fr}$ (is a front-running adversary) if it satisfies the following conditions: 
\begin{enumerate}
    \item Upon receiving a broadcast message by a party, it can maximally delay it by one round.
    \item Any broadcast message by the adversary is propagated through the network immediately. 
\end{enumerate}
Under such an adversary, there exist attacks discussed in~\cite{Eskandari-2019-frontRunning, Breidenbach-2018-HydraFR, Kalodner-2015-NamecoinFR} such as (1) \emph{displacement} attack and (2) \emph{insertion} attacks. For our purposes, however, the Definition~\ref{def:front-running} is of interest. 
\end{definition}

\subsection{Rational Protocol Design}
\label{ssec:rpd-prelims}

Rational Protocol Design~\cite{Garay-2013-RPD} (RPD) is the basis on which multiple security analysis models for PoW blockchain protocols is based, including~\cite{Badertscher-2018-BitcoinRPD, Badertscher-2021-Bitcoin51}. Our work motivates from the RPD and modifies it in context of PoW Blockchain protocols to a more practical model of Game Theoretic security analysis.

RPD models security as a game between two players, (1) the Protocol Descriptor (PD) which proposes the protocol that realizes the relaxed functionality and, (2) the Adversary, which chooses attack strategy once the PD has chosen a protocol. The Game $\mathcal{G}_{\mathcal{M}}$ is modelled as a two-player game with complete information and finite horizon (of two steps) -- a Stackelberg Game. RPD models deviating (adversaries) and non-deviating (altruistic) players. 

Badertscher et al.~\cite{Badertscher-2018-BitcoinRPD} modified RPD protocol curating to PoW blockchains. This protocol captured deviations by the protocol across rounds. However, this model didnot capture dynamics of the system such as variable mining difficulty which can be strategically manipulated by the adversary (as we see in Section~\ref{ssec:bz-attacks}), variable block-reward among other things. Further, the game $\mathcal{G}_{\mathcal{M}}$ models deviating (adversarial) and non-deviating (altruistic) players. They do not capture \emph{conditionally-deviating} (rational) players. We aim at introducing a Practical Rational Protocol Design model (pRPD Section~\ref{sec:pRPD}), which can model such setting.

\section{Attacks on PoW blockchains}
\label{sec:pow-attacks}
Previous models of security analysis such as~\cite{Garay-2013-RPD,Badertscher-2018-BitcoinRPD,b14} are good contributions. Still, their analysis is limited to a fixed difficulty, finite horizons, constant block reward, constant conversion rate, and a single mining round. In practical scenarios, these factors are dynamic, and an adversary can leverage them to launch attacks. This section discusses the attacks we have discovered on PoW-based blockchains that previous works fail to capture. We also discuss why previous works fail to capture these attacks.

We categorize these attacks into three categories, (i) \emph{Byzantine} adversary attacks -- the worst form of attack where a byzantine adversary with $\beta_{a} < \frac{1}{2}$ can single-handedly launch the attack. (ii) \emph{Rational-Byzantine} attacks -- where a byzantine adversary relies on rational agents to successfully launch an attack. (iii) \emph{Rational} attacks -- deviations of Rational Party from the original protocol $\Pi$. In these attacks, no byzantine adversary is involved. 

\subsection{Byzantine adversary attacks}
\label{ssec:bz-attacks}
We discovered that a byzantine adversary could launch a forking attack and double spend. Previously it was considered that if $\beta_{a} < \frac{1}{2}$, then the protocol is considered secure against such double spending attacks with a very high probability for reasonable block-confirmation time. However, we show that even for a very liberal block confirmation time, there exists an adversarial attack that can cause forks with very high probability even for $\beta_{a} < \frac{1}{2}$.

\subsubsection{Difficulty altering attack}
\label{sssec:daa}
The \daa\ takes place in two consecutive epochs $e_{i}$ and $e_{i+1}$. In epoch $e_{i}$, \ap\ slow down their apparent mining rate. This allows \ap\ to mine blocks with lower difficulty value on $C_{A}$ after difficulty recalculation in the epoch $e_{i+1}$ and overtakes $C_{H}$. Note that, in this attack \ap\ mines on $C_{A}$, whereas \hp\ and \rp\ mine on $C_{H}$.

\noindent \textbf{Attack Strategy.} In $e_{i}$, \ap\ forks the blockchain to form a private chain $C_{A}$ when $r_{1}$ fraction of $e_{i}$ is completed (i.e., $r_{1}\lambda$ blocks are mined). \ap\ creates blocks with timestamps such that the target recalculation leads to a very low difficulty for $C_{A}$ in the next epoch $e_{i+1}$. Consequently, it can  mine the blocks faster than \hp\ and \rp, and overcome $C_{H}$ to become the longest chain. $C_{A}$ overtakes $C_{H}$ when $r_{2} (\in (0,1])$ fraction of total blocks in the epoch $e_{i+1}$ (which is $r_{2}\lambda$ blocks) are mined. 

Notice that \ap\ does not need to mine the blocks slower. They have to put timestamps such that the blocks appear to be mined slower when made public. Further, broadcasted blocks must have timestamp $<$ broadcasted time. Thus, while \ap\ mines the blocks in epoch $e_{i}$ slower than \hp s (because $\beta_{a} < \frac{1}{2}$) due to the reduced difficulty of the private chain, it mines the remaining blocks of epoch $e_{i+1}$ faster than \hp\ and \rp. 

\begin{figure}[!ht]
\label{fig:daa}
\includegraphics[width=0.7\textwidth]{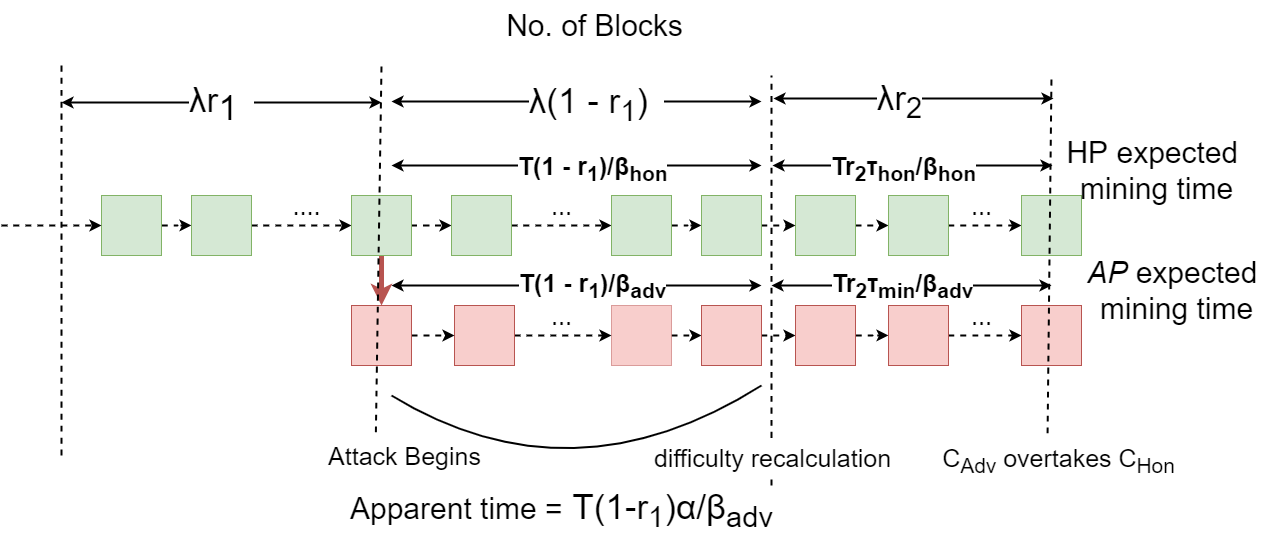}
\centering
\caption{Difficulty Altering Attack}
\end{figure}

\noindent \textbf{Analysis.} Let $T$ be the time to mine $\lambda$ blocks of epoch  $e_{i}$ if  all parties mine on $C_{H}$. If \daa\ is launched, let $\tau_{hon}$ be the difficulty adjustment for $C_H$ and $tau_{adv}$ for $C_A$. If \daa\, after difficulty-recalculation, in epoch $e_{i+1}$, $T\tau_{hon}$ is the time to mine all the blocks on chain $C_{H}$ and $T\tau_{adv}$ is the time to mine on chain $C_{A}$. 
Note that,  in this attack, $1-\beta_{a}$ computing power is mining on $C_{H}$ and $\beta_{a}$ on $C_{A}$. Hence, all the expected required times need to be normalized accordingly. 

The total time taken for $C_{A}$ to complete $r_{2}$ fraction of $e_{i+1}$ epoch should be $< C_{H}$, which gives us 
\begin{equation}\label{eq:daa-time}
    \frac{T\tau_{h}r_{2}}{1 - \beta_{a}} + \frac{T(1-r_{1})}{1 - \beta_{a}} > \frac{T(1-r_{1})}{\beta_{a}} + \frac{Tr_{2}\tau_{adv}}{\beta_{a}}
\end{equation}
In the equation above, the adversary slows down the apparent mining rate on $C_A$ by a factor of  $\alpha$ ($> 1$). 
Therefore, $\tau_{adv}$ and $\tau_{hon}$ are calculated as,
\begin{center}
$\tau_{adv} = max\Big(\frac{1}{r_{1} + \frac{\alpha(1-r_{1})}{\beta_{a}}},\tau_{min}\Big) \hspace{20pt} \tau_{hon} = max\Big(\frac{1}{r_{1} + \frac{(1-r_{1})}{1 - \beta_{a}}},\tau_{min}\Big)$
\end{center}

\begin{theorem}[\daa\ Attack]\label{thm:daa-attack}
When $\tau_{min}<\frac{1}{2}$, an adversary can fork a PoW blockchain using \daa\ attack w.p. $> 1- negl(\Theta\varepsilon)$ if $\beta_{a} \geq \underline{\beta_{a}}$.\\

\noindent (i) expected time to mine a block by any party is $\propto \Theta$, \\
(ii) $2\varepsilon$ is the difference in time between \ap\ and \rp\ mining the last block of the epoch $e_{i+1}$, and  \\
(iii) $\underline{\beta_{a}}= \frac{(3 + \tau_{min}) - \sqrt{(3 + \tau_{min})^{2} - 4(\tau_{min} + 1)}}{2}$

\end{theorem}

\begin{proof}
The proof is in three steps. In Step 1, we state what environment the adversary sets to maximize its utility. In Step 2, we determine the fraction of computing power required by the adversary to launch the attack with the overwhelming probability, which we quantify in Step 3. The complete proof is in Appendix~\ref{ssec:app-daa-attack}.
\end{proof}

\begin{corollary}[Bitcoin-\daa]\label{corr:bitcoin-daa}
Bitcoin is insecure against \daa\ Attack for $\beta_{a} > 0.4457$.
\end{corollary}
The result follows from putting the value of $\tau_{min} = \frac{1}{4}$, as used in Bitcoin. Previously, bitcoin was considered secure against forking, with a high probability for $\beta_{a} < \frac{1}{2}$, however, we thus show that forking is possible even for $0.4457 < \beta_{a} < \frac{1}{2}$. One might argue this attack reduces the currency's price because it is a compromise in security, and $\theta = e_{security}$ after this attack. Therefore, any profit the \ap\ gains in cryptocurrency is wasted. However, \ap\ can profit from holding short position for the coin. We explain this incentive manipulation in detail in Appendix~\ref{app:goldfinger}

\subsection{Rational-Byzantine attacks}
\label{ssec:rat-bz-attacks}
The rational-Byzantine attack is when an \ap\ launches an attack and relies on deviation from \rp\ for the attack to be successful. We discuss two attacks, (i) \pcf\ -- security threat, and (ii) \smb\ -- fairness threat to the protocol. 

\subsubsection{Quick Fork Attack}
\label{sssec:qf-attack}
In \pcf, the adversary creates a fork $k$($< \rho$) blocks before the latest block of the longest chain. If the \eo\ does not observe the fork as an attack, \rp s  are incentivized to mine on this forked chain for higher expected payoff. Since \hp\ continues to mine on $C_{H}$, the deviating parties collect a larger fraction of the reward. Attacks discussed in~\cite{Katz-2017-Whale1,Katz-2017-Whale2} are different from \pcf\ since the \pcf\ does not require abnormally large transactions to enable \rp\ to deviate. It is not observable by \eo, whereas the former attacks are easily observed by \eo\footnote{eg. transaction -- cc455ae816e6cdafdb58d54e35d4f46d860047458eacf1c7405dc634631c570d}.
\\
\begin{figure}[!th]
  \centering
  \includegraphics[width=0.7\textwidth]{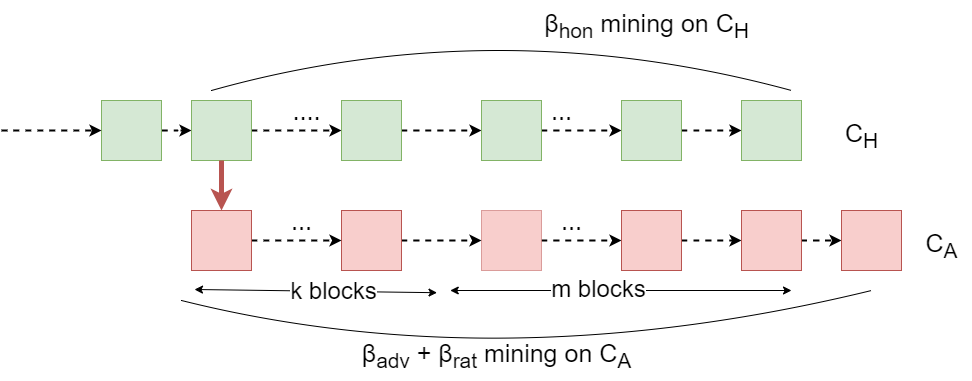}
  \caption{Successful Quick Fork Attack}
  \label{fig:quick-fork}
\end{figure}

\noindent \textbf{Attack Strategy.} 
Let $C_{H}$ be the honest chain, and $C_{A}$ be the forked chain. To launch this attack, the adversary has to ensure that at no time $C_{H}$ has a lead of $\rho$ or more blocks over $C_{A}$, where $\rho$ is \emph{tolerance factor} for \eo. This ensures that the fork is not considered an attack by the \eo, thereby incentivizing \rp\ to mine on the forked chain as they may grab more block rewards than mining on the $C_{H}$.

\begin{definition}[\pcf]
We say the adversary has launched \pcf\ successfully if (i) it forks the honest chain $C_H$ at $k<\rho$ blocks previous to the latest block on $C_H$ creating $C_A$, (ii) $C_A$ becomes the longest chain, and (iii) $C_{H}$ does not exceed $C_{A}$ by $>\rho$ blocks during the attack. 
\end{definition}

If $k \geq \rho$, \rp\ do not mine on the forked chain $C_A$, as $C_A$ is observed as an attack by the \eo\ and conversion rate $\theta(t)$ falls to $e_{security}$. Due to this, \rp s do not shift to the $C_{A}$, which therefore does not overtake $C_{H}$. Therefore, \pcf\ is possible only at $k <\rho$ blocks from the end of the longest chain.

\noindent \textbf{Analysis.} We show that the \pcf\ attack strategy is dominant over $\Pi$ for \ap\ and \rp.  We show conditions under which \pcf\ attack is possible with high probability. 
\begin{theorem}[\pcf\ Attack]\label{thm:pcf-attack}
Adversary successfully launches \pcf\ with probability of $\frac{n-1}{n}$ if
\begin{itemize}
    \myitem{$\eta > \frac{n - \beta_{h}n - 1}{n\beta_{h} - 1}$}
    \myitem{$\beta_{a} + \beta_{r} \geq \frac{1}{2}$}
    \myitem{$\beta_{h} > \frac{\chi}{r_{block}}$}
    \item{\small{$k < \rho - \varpi$}}    
\end{itemize}

\noindent Here, the cost incurred by the system on mining one block is $\chi_{1}$.  $r_{block}$ is the block-reward for the current phase $P_{i}$ and $\eta = \frac{r_{block}}{\chi_{1}}$, $\phi = \frac{\beta_{h}}{\beta_{a} + \beta_{r}}$ and $\varpi = log_{\phi}(\frac{1 + (n-1)\phi^{\rho}}{n})$
\end{theorem}
\begin{proof} In the proof, we compare payoff for \rp, and \ap\ in following and deviating from the protocol from $C_{H}$ state of the blockchain, from when the fork happens, till $C_{A}$ overtakes $C_{H}$. The proof goes into four steps. In Step 1, we compute the payoff on deviation; in Step 2, we calculate the payoff on following the protocol. In Step 3, we compare the results from Step 1 and Step 2 and show the conditions that make \pcf\ feasible. In Step 4, we argue about the probability which the attack is successful. The complete proof is provided in Appendix~\ref{app:pcf-attack-proof}.  
\end{proof}

\subsubsection{Selfish Mining with Bribing attack}
\label{sssec:smb-attack}

 Eyal and Sirer~\cite{Eyal-2014-Selfish}, showed that with computing power of $\frac{1}{4}$, (i.e., $\beta_{a}=0.25$ or higher), the attack \emph{selfish mining} (described in Section~\ref{sssec:selfish-mining-eyal}) is feasible. In this attack, \ap\ mines block without revealing to the network (called private chain) and announce its private chain at appropriate times (refer to ~\cite{Eyal-2014-Selfish} for more details), grabbing more rewards than its fair share, leading to $\theta(t)=e_{fairness}$. The authors assume that all miners are either in \ap\ (part of the pool launching a selfish mining attack) or \hp. In our model, miners are in either \ap, \rp, or \hp. For such realistic scenarios, we prove that the attack is possible even if $\beta_{a}<\frac{1}{4}$. The key intuition is that the adversary can bribe \rp s to switch to its chain ($C_A)$, thereby increasing the success probability for the attack. In summary, the $\frac{1}{4}$ bound in~\cite{Eyal-2014-Selfish} is a special case of a more general bound on $\beta_{a}$ (Theorem~\ref{thm:smb-attack}) for feasibility of Selfish Mining attack. 

 Therefore, our contribution is to show that if we consider the system to have \hp, \ap\ and \rp, then the bound on selfish-mining is actually more general if we introduce bribing in Selfish mining.


\noindent \textbf{Attack Strategy.} 
The attack model is similar to that discussed in~\cite{Eyal-2014-Selfish}. The only difference is that in each block that the \ap\ mines (privately), it includes a \emph{bribe transaction}\footnote{easy to grab a small reward whoever mines on the adversarial block.}. The modified protocol proposed in~\cite{Eyal-2014-Selfish} mandates the parties to choose one of the competing blocks randomly and mine on top of the chosen block. This choice is not known to other parties or the \eo. Thus, \rp\ can choose to mine on top of the Adversarial Block to collect bribes without \eo\ realizing the deviation.

\noindent \textbf{Analysis.} The total mining power mining on top of a non-adversarial block is $\frac{\beta_{h}}{2}$ (because each miner in \hp\ chooses randomly one of $C_{H}$ or $C_{A}$ to mine on), while on an adversarial block, a total of $\frac{\beta_{h}}{2}+\beta_{r}+\beta_{a}$ mining power is mining. Therefore this chain wins the race (block is mined) with a higher probability of $\frac{\beta_{h}}{2}+\beta_{r}+\beta_{a}$. The analysis leads to Theorem~\ref{thm:smb-attack}, which is proven in Appendix~\ref{app:smb-proof}.

\begin{theorem}[\smb\ Attack]
\label{thm:smb-attack}
Attack strategy \smb\ is dominant over the Honest protocol for 
\begin{equation}\label{eqn:smb-our}
    \beta_{a} > \underline{\beta}^{SMB} = \frac{\beta_{h}}{2\beta_{r} + 4\beta_{h}}
\end{equation}
Here $\underline{\beta}^{SMB}$ is the lower bound on $\beta_{a}$ for which \smb\ attack is profitable. 
\end{theorem}

\noindent From Theorem~\ref{thm:smb-attack}, we derive bound on $\beta_{a}$ for \smb\ attack. \\
\noindent\underline{Case 1 : (Fully Rational Setting)} Here, $\beta_{h} = 0$ and therefore $\gamma = 1$. The attack is profitable for $\beta_{a} > 0$.\\
\noindent\underline{Case 2: (Mixed Setting)} In this setting $\beta_{h} > 0$, therefore, the bound when this attack is feasible for the attacker becomes $\beta_{a} > \underline{\beta}^{SMB}$. 

By trivial analysis, we see that $\underline{\beta}^{SMB} \leq \frac{1}{4}$ (equality when $\beta_{r}=0$). However, the actual $\underline{\beta}^{SMB}$ can be much smaller depending on $\beta_{r}$. The gist of the analysis is that selfish mining is possible with much less computing power than the previously known bound of $25\%$ in the presence of rational parties.

\subsection{Rational attacks}
\label{ssec:rat-attacks}
Rational Attacks are deviations from the original protocol $\Pi$ or some sub-routine of $\Pi$. In this case, each rational agent deviates from the protocol, and no adversary is involved. We discuss one such deviation from the \emph{Gossip} sub-protocol of $\Pi$, the transaction withholding attack. 

\subsubsection{Transaction Withholding attack ($\sigma_{tw}$}
\label{sssec:tx-withold}

When a miner receives a transaction, they add it to the mempool and share it with their peers, which is the expected behavior of Gossip protocol $\Pi_{Gossip}$. 
In Block-Reward Model (BRM) \rp\ has a negligible incentive to deviate from $\Pi_{Gossip}$. In the Transaction Fee Only Model, such deviations are incentivized, as shown by Theorem~6 in~\cite{Badertscher-2018-BitcoinRPD}. We show one such deviation -- transaction withholding attack (represented as $\sigma_{tw}$) which dominates $\Pi_{Gossip}$ as shown in Lemma~\ref{lemma:gossip-subopt}. 

\begin{lemma}
\label{lemma:gossip-subopt}
In TFOM, for any rational party, the following $\sigma_{tw}$ strictly dominates $\Pi_{Gossip}$.
\end{lemma}
\begin{proof}
We first calculate the payoff for a \rp\ in the following $\Pi_{Gossip}$ to prove this. We then calculate the payoff on the following $\sigma_{tw}$ and show that the latter payoff is strictly greater than the former. We provide the complete proof in Appendix~\ref{app:gossip-subopt-proof}.
\end{proof}
\noindent This deviation poses one (or both) of two possible threats:
\begin{itemize}
    \item[\ding{64}] \underline{Low Throughput}: The chance of a single \rp\ mining a block is minimal; therefore, the transactions take a long time to get accepted in the blockchain, thus reducing the throughput. 
    \item[\ding{64}] \underline{Centralization}: The protocol becomes centralized if the transaction proposer sends the transaction to a large mining pool in the hope of getting the transaction published quickly.
\end{itemize}

\subsection*{Where do previous frameworks fail?}
We discussed four possible attacks in PoW blockchains. We shall discuss why previous models (most specifically Rational Protocol Design (RPD)) could not capture these deviations. Three properties were not captured by previous works and are discussed below.

\subsubsection{Agent Modelling} Some miners could be rational, i.e., want to maximize their utilities without disrupting the protocol. In~\cite{Badertscher-2018-BitcoinRPD}, \pd's utility comprises of honest miner's utility and a very high negative payoff for events such as forking ($exp(polylog(\kappa))$). It fails to capture the objective of the \pd\ -- maximize the difference between parties that follow protocol and parties that deviate. Thus, RPD could not capture deviations discussed in Section~\ref{ssec:rat-bz-attacks}.

\subsubsection{Externalities} In RPD, the conversion rate of the underlying crypto-currency to fiat currency is considered constant. However, the loss of utility due to security attacks should be calculated through a change in \emph{externality}, which affects $\theta(t)$. Modeling such externality allows quantifying the loss in utility for \ap\ and \rp\ on deviating from the protocol. 

\subsubsection{System Dynamics} RPD and its derivatives do not account for (i) block rewards changing over time and (ii) variable difficulty of mining. Because of this, previous models could not capture attacks discussed in Section~\ref{ssec:bz-attacks}.

\section{\prpd \ for Blockchains}
\label{sec:pRPD}
In this section, we introduce the Practical Rational Protocol Design(\prpd]), an improved model for the game-theoretic security analysis of PoW blockchains. pRPD models blockchain as a two-player Stackelberg game between Protocol Descriptor and Adversary. 

\subsection{Players}
There are two players in our modeling of the protocol as a game: (i) \pd\ and (ii) \ap. 
\begin{itemize}
    \item \textbf{Protocol Descriptor (PD)}: The protocol descriptor is the player who selects the protocol $\Pi$ which will be considered as the honest protocol. PD must ensure that (i) The relaxed ideal-functionality $\mathcal{G}_{weak-ledger}$ is realized by this functionality, and (ii) This protocol gives them the maximum utility.
    \item \textbf{Adversarial Party (AP):} We exploit the notation to use \ap\ to refer to both the set of adversarial miners and the adversary who is part of the two-player game. This is because the set of adversarial miners is controlled by a PPT ITM $\mathcal{A}$, which carries out the adversarial strategy. This $\mathcal{A}$, which decides the adversarial strategy given a protocol $\Pi$, is the \ap\ in this game.  
\end{itemize}

\subsection{Attack Model} 
\label{ssec:attack-model}
\noindent As defined in~\cite{Garay-2013-RPD, Badertscher-2018-BitcoinRPD}, we need to define an attack model for game-theoretic analysis of the protocol's security. The attack model is parameterized by the tuple $(\mathcal{F},<\mathcal{F}>,v_{A},v_{D})$. In this, $\mathcal{F}$ is the ideal functionality that the protocol wishes to realize in the real world. $<\mathcal{F}>$ is the relaxation of the ideal functionality. In the case of PoW Blockchain protocols, $\mathcal{G}_{ledger}$ is the ideal functionality, and the $\mathcal{G}_{weak-ledger}$ functionality as its relaxation.

Further, $v$ are mappings $v:\mathcal{S}\times\mathcal{Z}\longrightarrow\mathbf{R}$ from the simulator, simulating attack strategy $\mathcal{A}$ on protocol $\Pi$, in environment $\mathcal{Z}$ to real-valued payoff. The utility is defined as the expectation over this payoff function. In our case, the three types of parties \hp, \rp\, and \ap\ have payoff vectors $v_{H}$,$v_{R}$, and $v_{A}$ respectively. Therefore, our attack-model is $$\mathcal{M} = (\mathcal{G}_{ledger},\mathcal{G}_{weak-ledger},v_{A},v_{R},v_{H})$$

\subsection{Game}
\label{sssec:game-des}
\noindent Typically, a blockchain system consists of two types of participants 
(i) protocol descriptor and (ii) miners -- we call each type of miner as \emph{parties} (already described in Section~\ref{sssec:parties}). Additionally, we assume an \emph{external observer} who is not part of the game. The game progresses as follows:
\begin{itemize}
    \item[\ding{112}] The \pd\ defines the protocol. Its role is to choose the best protocol $\Pi$, which maximizes its utility (and, by design, of the players who follow $\Pi$).
    \item[\ding{112}] \ap\ observes $\Pi$ and chooses an attack strategy implemented by any ITM $\mathcal{A} \in \mathcal{C}_{A}$ for the choosen $\Pi$.
\end{itemize}

\noindent We model the interaction of the adversary parties with the system as a two-player Stackelberg game between \pd\ and the adversary where the leader of the game is \pd, and the follower is \ap. In the system, there is \eo\ who determines the conversation rate $\theta(t)$ -- the price of one unit of underlying cryptocurrency to a fiat currency at round $t$. Our game. $\mathcal{G}_{\mathcal{M}}$, is defined over attack-model $\mathcal{M} = (\mathcal{G}_{ledger}, \mathcal{G}_{weak-ledger}, v_{H}, v_{R}, v_{A})$. 
The \rp\ and \hp\ are not part of the Stackelberg game because for \hp\ strategy is fixed $\Pi$, when \pd\ moves, and for \rp\ the most optimal out of the possible deviations is fixed after \ap\ selects their strategy.

\subsubsection{Security Definitions}
\label{sssec:security-defs}
Similar to~\cite{Badertscher-2018-BitcoinRPD, Badertscher-2021-Bitcoin51} say the `best possible behavior' that is practically achievable for the adversary is a semi-honest front-running strategy. The front-running semi-honest adversary $A_{fr}$ is a subset of front-running adversary $A_{fr}^{*}$.
\begin{definition}[Front-Running Semi-Honest Adversary (Def. 2,~\cite{Badertscher-2018-BitcoinRPD})]\label{def:semihonest-front-running}
An adversary $\mathcal{A}$ which is in the set of adversaries $\mathcal{A}_{fr}$ is said to be semi-honest, front-running if
\begin{itemize}
    \item upon activation, the adversary corrupts miners and follows $\Pi$ (honest protocol). 
    \item any message that the adversary wants to broadcast, it does so immediately
    \item if a non-adversarial party wants to broadcast the message, the adversary maximally delays that message by one round.  
\end{itemize}
\end{definition}
Now that we have defined the behavior of the adversary we want to achieve, we define the strongest achievable security guarantee for a protocol: strong attack-payoff security. We motivate the definition of \textit{strong attack-payoff security} from~\cite{Badertscher-2018-BitcoinRPD}. 
\begin{definition}[Strong Attack-Payoff Secure (Def. 3,~\cite{Badertscher-2018-BitcoinRPD})]\label{def:strong-aps}
A protocol $\Pi$ is \textit{strongly attack-payoff secure} under attack-model $\mathcal{M}$ if for some adversary in the set of semi-honest, front-running adversary $A \in \mathcal{A}_{fr}$, the attacker playing $A$ is approximate best response strategy. That means, $\forall A_{2} \in \text{PPT ITM}$ and $A_{1} \in \mathcal{A}_{fr}$, $$U_{A}(\Pi,A_{2}) \leq U_{A}(\Pi,A_{1}) + negl(\kappa)$$ 
\end{definition}


\subsection{Utility}
\label{ssec:utility-model}
For calculating the utility, we define random variables for the payoff for each $\mathcal{A},\mathcal{R}$, and $\mathcal{H}$. The random variable $v_{\mathcal{X}}$ for $\mathcal{X} \in \{\mathcal{A},\mathcal{R},\mathcal{H}\}$ is defined over environment $\mathcal{Z}$. On the lines of the Universal Composability paradigm, since the game is a Stackelberg Game, suppose that the \pd\ chooses strategy $\Pi$, and the \ap\ chooses $\mathcal{A} = A(\Pi)$. In that case, let $\mathcal{S}$ be an ideal-world simulator for $\mathcal{A}(\Pi)$, where $\mathcal{S}$ attacks on $<\mathcal{F}>$. The set of all such simulators is $\mathcal{C}_{A}$. 

\subsubsection{Objective of the Model}
\label{sssec:objective-utility}
The \ap's objective is to achieve more payoff for itself. Further, we observe that, for \ap\ and  \eo\ (both PPT ITMs), the \hp\ and \rp\ are indistinguishable from each other. Thus, the objective of the protocol descriptor is to achieve more payoff for \hp\ and \rp\ and less for the \ap. Notice that it is at this point that the utility model differs from most of the previous works. It also captures the dynamic nature of the protocol by considering payoff across different rounds with changing protocol parameters. 

Previous utility models captured only deviations which benefit the adversary. However, \ap\ can gain higher utility in the long run by reducing the payoff of \hp. We elaborate on how this is possible in Appendix~\ref{app:difference-utility}. The payoff of \pd\ is thus the difference between the payoff of non-deviating parties \hp\ and \rp\ (in view of \eo) and deviating party \ap.
Another distinction from RPD is how externalities are modeled in the payoff of the miners. In our work, we account for critical security threats and moderate fairness threats through externality (reflected by the conversion rate $\theta(t)$). This term decreases both \ap\ and \pd\ payoff. However, \ap\ payoff can increase if they are holding short position against the cryptocurrency. To model such situations, while calculating $U_A$, we add an extra term in $v_A$, which is inversely proportional to the currency's conversion rate. 
\subsubsection{Utility Model}
\label{sssec:our-utility}

Consider the attack model as described in Section~\ref{ssec:attack-model}. If an attack $A(\Pi)$ is defined on a chosen protocol $\Pi$, let $C_{\mathcal{A}}$ be the set of simulators, which simulate the attack on the relaxed, ideal functionality $<\mathcal{G}_{weak-ledger}>$. Given $\mathcal{S} \in C_{\mathcal{A}}$ and environment $\mathcal{Z}$,  $v_{H}, v_{A}, v_{R}$ are expected payoff of \hp, \ap, and \rp. We find the normalized payoff of a single miner as $\frac{v_{H} + v_{R}}{n(t)(\beta_{Hon} + \beta_{Rat})}$ for single honest (non-deviating for \eo) miner and $\frac{v_{A}}{n(t)\beta_{Adv}}$ for single \ap\ (deviating w.r.t. \eo). If we multiply the utility model with a positive constant, we use the fact that the players' best strategies do not change and drop $n(t)$ -- the number of miners in the system at round $t$. 

For \ap, we minimize over this set $C_{\mathcal{A}}$ for the choice of simulator $\mathcal{S}$. This is because $C_{\mathcal{A}}$ contains all the simulators which can launch the attack. Many of these may invoke additional events unrelated to the attack. However, the purpose of \ap\ is to force the simulator to invoke the attack in the ideal world. Hence, the most closely related payoff is of the simulator that ``just'' simulates the attack in the ideal world. Hence, we minimize over the set of all simulators, $\mathcal{C_{A}}$ for \ap. 
For \pd's utility, which is a function of the expected payoff of \hp, \rp\, and \ap, we consider the worst environment ($\mathcal{Z}$), and for \ap's utility, the best environment. In summary, utilities for a given adversarial strategy $A(\Pi)$ for \pd\ strategy $\Pi$ is as follows.  \\
\begin{equation}
U^{\Pi,<\mathcal{F}>}_{\mathcal{D}}(A) = \min_{\mathcal{Z}\in ITM}\Big\{\min_{\mathcal{S} \in C_{\mathcal{A}}}\Big\{\frac{v_{\mathcal{H}} + v_{\mathcal{R}}}{\beta_{Hon} + \beta_{Rat}} - \frac{v_{A}}{\beta_{Adv}}\Big\}\Big\}
\end{equation}
\begin{equation}
U^{\Pi,<\mathcal{F}>}_{\mathcal{A}}(A) = \max_{\mathcal{Z}\in ITM}\Big\{\min_{\mathcal{S}\in C_{\mathcal{A}}}\Big\{\frac{v_{A}}{\beta_{Adv}}\Big\}\Big\}
\end{equation}

Our adversary is an \emph{$\mathcal{M}-$maximizing adversary}. This means that given a protocol $\Pi$ chosen by the \pd, the adversary chooses the best response attack $A$, which maximizes their utility function $U_{A}(\cdot)$. 
\begin{definition}\label{def:m-maximizing-adv}
An adversary is $\mathcal{M}-$maximizing adversary if given protocol $\Pi$ which realizes functionality $<\mathcal{F}>$, they choose strategy such that for their utility function $U^{\Pi,<\mathcal{F}>}_{\mathcal{A}}(\cdot)$ is maximized. 
\begin{equation*}
    U_{\mathcal{A}}^{\Pi,<\mathcal{F}>} = \max_{A \in ITM} U_{\mathcal{A}}^{\Pi,<\mathcal{F}>}(A)
\end{equation*}
\end{definition}

\subsubsection{Advantages of Our Utility Model}
\label{sssec:util-adv}

The advantages of our utility model are as follows:
\begin{enumerate}[leftmargin=*]
    \item It models the \emph{externality} more flexibly, allowing an \eo\ who can observe certain types of deviations as attacks and reduce the $\theta(t)$ correspondingly. This also allows us to model market responses differently to different type of attacks.
    \item It can represent \emph{variable block reward}. Further, it does not restrict to a specific type of reward model\footnote{most of the previous works have stuck to constant block-reward}, 
    but a general series that can converge or diverge. 
    \item It captures \emph{variable difficulty} because the probability of mining in each round is different from each other. 
    \item The \emph{utility} of the protocol descriptor does not just try to increase the payoff of \hp\ but also decreases the difference between \hp\ and \ap\ utility. This is better than previous models because it does not allow high utility for such attacks, which increases the \hp\ utility but increases the \ap\ utility even further (possibly due to external payoffs). 
\end{enumerate}

\section{Detering Attacks}
\label{sec:sol}
In this section, we propose modifications to the original PoW blockchain protocol that helps us tackle the attacks discussed in Section~\ref{sec:pow-attacks}. 

\subsection{Difficulty Altering Attack}
\label{ssec:sol-daa}
From Theorem~\ref{thm:daa-attack}, it is clear that we can overcome possibility of \daa\ attack if $\underline{\beta_{adv}}$ is at least $\frac{1}{2}$. If $\beta_{adv}>\underline{\beta_{adv}}=0.5$, then no PoW blockchain is secure. In this subsection, as a corollary to Theorem~\ref{thm:daa-attack}, we show that we can achieve this if we appropriately set $\tau_{min}$. 

\begin{corollary}\label{corr:daa-sec}
A PoW blockchain with $\beta_{adv} < \frac{1}{2}$ is secure against \daa\ Attack if $\tau_{min} \geq \frac{1}{2}$
\end{corollary}
\begin{proof}
From the proof of Theorem~\ref{thm:daa-attack}, lower bound on $\beta_{adv}$ to launch \daa\ Attack is 
$\underline{\beta_{adv}}= \frac{(3 + \tau_{min}) - \sqrt{(3 + \tau_{min})^{2} - 4(\tau_{min} + 1)}}{2}$. We want $\underline{\beta_{adv}}\geq\frac{1}{2}$. With simple algebra, one can argue that $\underline{\beta_{adv}} \geq 0.5$ if 
$ \tau_{min} \geq \frac{1}{2}$.\footnote{we do not discuss $\beta_{adv} > \frac{1}{2}$ because forking is possible in that case by $51\%$ attack.}
\end{proof}

\subsection{\pcf\ Attack}
\label{ssec:sol-pcf}
To defend against \pcf\ attack, we exploit that \rp\ and \hp\ are indistinguishable for \ap. For \rp\ to mine on adversarial chain ($C_A$) with the help of \rp, \ap\ makes $C_A$ public making it visible to \hp s too. We propose that all parties (non-deviating) are allowed to add \emph{Proof-of-Invalidity} (PoI) to any chain shorter than the chain on which they are mining. With POI, we can prevent \pcf\ attack. 

\noindent\textbf{\texttt{PC-MOD} Solution}
We propose that blockchain protocols allow party $par$ mining on chain $C_{par}$ at height $a_{par}$ to add a block containing POI on a forked chain ($C_{A}$) if it observes $C_{A}$ is of height $\leq a_{par} - k_{th}$. In this, $k_{th}$ is the \emph{threshold gap}, which is defined below.
\begin{definition}[Threshold Gap]
Threshold Gap ($k_{th}$) is the difference in the height of the longest chain($C_{H}$) and forked chain ($C_A$) such that $\beta_{rat}+\beta_{adv}$ mining on $C_{A}$ can overtake $C_H$ w.p. $\geq 1 - \mu$. 
\begin{equation}\label{eqn:kth}
    k_{th} = \rho - log_{\phi}(\mu + \phi^{\rho}(1 - \mu))
\end{equation}
Here, $\phi = \frac{\beta_{hon}}{1 - \beta_{hon}}$ and $\rho$ is the tolerance factor of the \eo. This relation is derived by following a similar argument as in Step 4 of Theorem~\ref{thm:pcf-attack} and results from Section 4.5.1 of~\cite{Ross-Probability}
\end{definition}
\noindent We want PoI should satisfy two conditions. (1) A block containing that POI is indistinguishable from any other block; otherwise, the \rp\ and \ap\ ignore that block and mine on top of its parent block, and (2) Adversary should not be able to add POI on the honest, longest chain and invalidate that chain. We construct PoI from the definition below, which satisfies the two requirements. 

\begin{definition}[Proof of Invalidity]\label{def:poi}

The PoI \emph{transaction} is published on the forked chain to prove its invalidity. It is constructed as:
\begin{itemize}
    \item PoI consists of a string $h$ which is the hash $H(H_{a_{par}} || m_{secret})$\footnote{here $H$ is the hash-function used in PoW blockchain.}. Here $H_{a_{par}}$ is the hash of the block at height $\geq a_{par}$ on the $C_{H}$, and $m_{secret}$ is a secret string chosen by the proposer of PoI.
    \item Since $h$ is an arbitrary string, this transaction is indistinguishable from any other transaction.
    \item Since the proposer of the invalidity transaction is \hp, they can invalidate the chain $C_{A}$ if it overtakes the $C_{H}$ as the longest chain by revealing the $m_{secret}$. If the PoI is added to $C_{A}$ at height $\leq a_{par} - k_{th}$, the PoI is considered valid.
\end{itemize}
\end{definition}

\begin{claim}\label{claim:pro-claim}
The probability of \hp\ being successful in adding POI on $C_A$ is $\geq 1-e^{-\beta_{hon}\cdot k_{th}}$.
\end{claim}
\begin{proof}
Initially, $C_{A}$ trails behind $C_{H}$ by $k_{th}$ blocks. \hp\ can add POI in these $k_{th}$ blocks only because the height at which POI is present should be less than the height of the block whose hash it contains (which is at height $a_{hon}$). Therefore, if an adversary mines a block among these $k_{th}$ blocks, they can successfully add POI to that block. Let $E_{mit}$ be a chance that \hp\ successfully mines a block in these $k_{th}$ blocks. Using the inequality $e^{-x} > 1 - x$ we get 
$P[E_{mit}] = 1 - (1 - \beta_{hon})^{k_{th}} > 1 - e^{-\beta_{hon}k_{th}}$
\end{proof}

Observe that the above probability only accounts for \hp s adding POI via mining a block. In practice, \hp\ can broadcast the POI transaction, and all parties mining on $C_A$ add the transaction as it is indifferent from any other transaction. Thus, the actual probability is higher than in Claim~\ref{claim:pro-claim}. 

\begin{theorem} PoW blockchain protocol with \textbf{\texttt{PC-MOD}}\\ 
(i) it is an equilibrium for \rp s to mine on the longest chain and not to shift to $C_A$, the forked chain in \pcf\ attack.
(ii) Protocol is secure against \pcf\ attack for $\beta_{adv} < \frac{1}{2}$ with high probability.
\end{theorem}
 
\begin{proof}
The \texttt{PC-MOD} modification mandates \hp s to mine on the $C_{A}$ unless their POI transaction is included in one of the blocks in the chain. This makes the deviation to launch (and join) \pcf\ Attack disincentivized due to three reasons:
\begin{itemize}
    \item[1] All $\beta_{adv} + \beta_{rat} + \beta_{hon}$ parties mine on $C_{A}$, so any advantage that the \rp\ or \ap\ might have gotten due to increased share of block-reward is now not present. 
    \item[2] \rp\ is disincentivized to mine on $C_A$ as (i) the block reward is not higher than mining on the longest chain $C_H$, and (ii) if \rp\ shifts to mining on $C_A$, the mining cost on $C_H$ between on $C_H-C_H^{\lfloor k^{th}}$  is wasted, implying lesser utility than mining on $C_H$.
    \item[3] If such an attack still takes place, there is always the risk of \hp\ mining a block or a valid POI transaction (indistinguishable from other transactions) is included in the $C_{A}$, which leads to $\theta(t) = e_{security}$. This happens with probability $> (1 - (1 - \beta_{hon})^{k_{th}})$ thus disincentivizing the \rp\ from participating in the attack.
\end{itemize}
Thus, for $\beta_{adv} < \frac{1}{2}$, the attack happens only if $\beta_{adv}$ can fork the chain by itself, which is possible with negligible probability. 
\end{proof}

\subsection{\smb\ Attack}


We show a rather pessimistic result in case of \smb\ attack. This result shows that it is impossible for a protocol which realizes the ledger functionality $\mathcal{G}_{weak-ledger}$ to be resilient to \smb\ attack. 

\begin{theorem}[Selfish-Mining Impossibility]\label{thm:smb-impossibility}
    For any PoW-Blockchain which UC-realizes  $\mathcal{G}_{weak-ledger}$, the protocol can't be \emph{strongly attack-payoff secure} because \smb\ is always possible if front-running is possible. 
\end{theorem}
\begin{proof}The proof follows as a direct result of the Lemma~\ref{lemma:smb-impossibility-lemma}, which is stated below. If for every protocol, each honest execution has an indistinguishable \smb\ counterpart, then for every such protocol, \smb\ attack is possible. Note that the proof, as standard in the literature~\cite{Badertscher-2018-BitcoinRPD,Badertscher-2021-Bitcoin51},  inherently assumes that the best response $A$ for any adversary is front running, i.e., $A\in\mathcal{A}_{fr}$.
\end{proof}
\begin{lemma}\label{lemma:smb-impossibility-lemma}
    For a protocol $\Pi$, there exists an environment $Z_{1}$ and a simulator for front-running adversary $S_{1} \in \mathcal{A}_{fr}$ and a corresponding environment $Z_{2}$ and an simulator for adversary using \smb\ $S_{2} \in \mathcal{A}_{SMB}$ such that for any PPT observer, execution $(S_{1},Z_{1})$ and $(S_{2},Z_{2})$ are indistinguishable.
\end{lemma}
\begin{proof}
    Proof is provided in Appendix~\ref{app:smb-impossibility-lemma}
\end{proof}
\subsection{\txincl\ Protocol}
\label{ssec:txIncl}
To resolve the Transaction withholding attack
(Section~\ref{ssec:rat-attacks}), we propose a modification in the form of an additional sub-protocol over the $\Pi_{gossip}$. This sub-protocol ($\Pi_{Tx-Inclusion}$) is a filter by which each miner can add only transactions satisfying a certain condition in the current block. With this modification, we can argue that \rp's gain in utility by withholding transaction is negligible, implying that following $\Pi_{gossip}$ is approximate Nash-equilibrium over the transaction withholding deviation. 
\subsubsection{Proposed Modification} 
A \rp\ mining a block can add only those transactions in the block which satisfy condition \textbf{C1}. 

\noindent \textbf{C1:} A transaction $tx$ satisfies \textbf{C1} given the block and coin-base address $dest$ (similar to Pay2PubHash in Bitcoin~\cite{pay2pub-wiki}) with parent block hash  $\mathfrak{H}$ if the last $l$ bits of $\mathbf{H}(tx,\mathfrak{H})$ and $\mathbf{H}(dest,\mathfrak{H})$ are same. If coin-base is a script (ex. Pay2ScriptHash in Bitcoin~\cite{pay2script-wiki}), $dest$ is the script hash. 
\begin{figure}[]
\centering
  \begin{tcolorbox}
\begin{center}
    $\Pi_{Tx-Inclusion}$
\end{center}

\textbf{Input}: $dest$\footnote{miner's destination address}, $tx$\footnote{transaction},$\mathfrak{H}$\footnote{hash of parent block}\\
\emph{if} at last $l$ bits $\mathbf{H}(tx,\mathfrak{H}) = \mathbf{H}(dest,\mathfrak{H})$ \emph{return} True\\
\emph{else} \emph{return} False
\end{tcolorbox}
\end{figure}

To add as many transactions as possible, \hp s may need to maintain multiple keys for which we can use PKI Trees (\cite{b9}) which takes logarithmic space for key storage. With these modifications, theprobability of a party mining a block and simultaneously including the withheld transaction is reduced because of one of two reasons:    
\begin{itemize}
    \item[1] If the party randomly selects an address and spends all the computing power on PoW for mining the block, there is a $\frac{1}{2^{l}}$ chance of that transaction being valid to be in the block. 
    \item[2] If the party spends some of its mining power on finding a favorable address mapping, then the number of queries they can perform for PoW reduces, thereby reducing their probability of mining a block. Also, the address mapping created by the party is not useful for the next round. 
\end{itemize}
\begin{lemma}\label{lemma:gossip-sol}
If \txincl\ is followed, \gossip is $\epsilon_{G}-$Nash Equilibria, for $\epsilon_{G} = tx\cdot O(2^{-l})$. Here, $tx$ is the cumulative fee from the transaction sent to the party 
 and is poly in $l$.
\end{lemma}
In summary, on following \txincl, \tw\ attack gives no significant payoff as shown in Lemma~\ref{lemma:gossip-sol}. The proof follows by computing the difference in the payoff of following and deviating from $\Pi_{gossip}$, which is $\leq \epsilon_{G}$. We provide the calculation in Appendix~\ref{ssec:app-txincl-proof}.


\section{PRAGTHOS  \& Theoretical Analysis}
\label{sec:mod-results}

We have discovered multiple attacks on PoW Blockchains (which also exist in Bitcoin). The previous game-theoretic analysis~\cite{Badertscher-2018-BitcoinRPD,Garay-2013-RPD,b14, Judmayer-2021-AIM} primarily focused on static population and horizon in which block-rewards and difficulty are constant. Our analysis framework, proposed in Section~\ref{sec:pRPD}, is very general and could discover the before-mentioned attacks (Section~\ref{sec:pow-attacks}). In Section~\ref{sec:sol}, we proposed novel solutions to these attacks by (1) Proposing additional sub-protocols in the PoW blockchain or (2) Specifying hyper-parameter values. With these modifications, we abstract out a new framework for PoW blockchain protocols. We call it  \proName, -- Practical Rational Game Theoretically Secure. It also is a conjunction of words \emph{`Pragmatic'} meaning logical (rational), and \emph{`Ethos'}, which roughly translates to character, describing the Rational Characteristic of the users of the protocol. In this section, we first summarize \proName, and then (Section~\ref{ssec:results}) provide its security analysis. 

\subsection{Modification to PoW Blockchain}
\label{ssec:mod}

\noindent \textbf{PoW Framework} As mentioned in Section~\ref{ssec:pow-protocol}, in a PoW blockchain, parties mine a block by solving a cryptographic puzzle. The puzzle encompasses finding a random nonce along with the merkle root of transaction data, header data is fed to hash again, and the final hash should be less than a certain target determined by difficulty recalculation at the start of each epoch. The parties are expected to collect all transactions they hear and adjust difficulty at the end of the epoch to maintain the average duration between two blocks as same as possible. The ratio of the previous difficulty and the new difficulty must be $\in [\tau_{min},\tau_{max}]$. The block rewards change by a factor $\vartheta$ across phases. Let the block-reward scheme followed by the protocol be given as $g(0),g(1),\ldots$ where $g(i) = r_{block}(0)\cdot\vartheta(i)$.\footnote{this relation can also be written as $r_{block}^{new} = \varpi(r_{block}^{old}, i) = r_{block}^{old}\frac{\vartheta(i)}{\vartheta(i-1)}$} denotes the block reward in Phase $i$. For bitcoin, $\vartheta(i) = \frac{1}{2^{i}}$.

When the sequence $<\vartheta>$ is converging (i.e. $\sum_{i=0}^{L} \vartheta(i)$ is finite for all $L$), the underlying crypto-currency is called \emph{deflationary}; otherwise we call it \emph{inflationary}. 

\begin{definition}[Inlfationary Crypto-Currency]
\label{defn:inflationary}
We say, a PoW crypto-currency is \emph{inflationary} if block-rewards update according to $r_{block}^{new} = \varpi(r_{block}^{old}, i) = r_{block}^{old}\frac{\vartheta(i)}{\vartheta(i-1)}$, and $<\vartheta>$ 
is diverging\footnote{diverging $\Rightarrow\lim_{L\rightarrow\infty} \sum_{i=0}^{L} \vartheta(i) \longrightarrow \infty$}.
\end{definition} 

With the modifications stated in Fig.~\ref{fig:pragthos} to PoW protocols, \proName\ is strongly attack-payoff secure if $\beta_{adv} < \frac{1}{2}$ and ensures fairness (against \smb\ attack) if $\beta_{adv} < \underline{\beta}^{SMB}$.
\begin{figure}[!th]
\begin{tcolorbox}[]
In \proName, the PoW blockchain undergoes the following modifications.
\begin{itemize}
    \item[\ding{111}] \emph{\texttt{PC-MOD}.} All parties are expected to add POI (Definition~\ref{def:poi}) if they observe a fork that is at least $k_{th}$(Eq.~\ref{eqn:kth}) block behind their current chain. 
    \item[\ding{111}] \emph{Difficulty Adjustment.} To protect against \daa\ Attack, the parties are expected to use $\tau_{min} = \frac{1}{2}$ while updating difficulty at the end of each epoch.
    \item[\ding{111}] \emph{For Adding Transactions.} For collecting transactions in a block, it follows $\Pi_{Tx-inclusion}$.
\end{itemize}
\end{tcolorbox}
\caption{Pragthos Framework}
\label{fig:pragthos}
\end{figure}

First, we need to assume that if every miner is honest, the reward structure is such that mining is profitable, compensating the costs incurred. We call it \emph{All-honest-profitability}. This condition ensures for all \hp\ the protocol is \emph{Individually-Rational}\footnote{Individual-rationality means payoff from participating in the protocol is $\geq$ the payoff from abstaining from participating.}. Note that, we are not assuming $\beta_H=1$ in the analysis.

\begin{definition}[All-honest-profitability]\label{def:ahp}
We say a PoW blockchain block-reward scheme satisfies \emph{All-honest-profitability} at round $t$ if for a system where $\beta_{H}=1$ we have $ \theta(t)r_{block}(t)p_{H} > \chi(t)$. Here, $p_{H}$ is the probability of a single miner mining a block in round $t$.  
\end{definition}
\subsection{Results}
\label{ssec:results}

PoW blockchains can be forked by \ap\ having majority computing power (through $51\%$ attack), due to which mining need not be profitable for \hp. Thus, we assume that $\beta_{adv} \leq \frac{1}{2}$. However, as indicated in Section~\ref{ssec:bz-attacks}, even with this, in a typical PoW, blockchain is susceptible to attacks which might lead to $\theta(t) = e_{security}$, making mining not profitable. 

\subsubsection{Strong Attack-Payoff Security for Inflationary Currency}
\label{sssec:results-inflation}
In this section, we analyze and prove in Theorem~\ref{thm:spas} \proName\ is strongly attack-payoff secure (Definition~\ref{def:strong-aps}) under an inflationary block-reward scheme (sufficiency condition). We further prove that such inflation in \proName\ is \emph{necessary} for any PoW blockchain protocol to be strongly attack-payoff secure (Theorem~\ref{thm:dra}). 
\begin{theorem}[Strong attack-payoff Security -- Sufficiency]\label{thm:spas}
\proName\ is \emph{strongly attack-payoff secure} under $\beta_{adv} < \frac{1}{2}$ if reward scheme is inflationary and satisfies All-honest-profitability.
\end{theorem}
\begin{proof}
To prove the result, we leverage the UC framework, originally developed by Canetti~\cite{Canetti-2001-UC}, further illustrated for blockchains by Badertscher et al.~\cite{Badertscher-2017-BitcoinUC}. We briefed it in Section~\ref{ssec:uc-prelims}. With this, the proof directly follows from Lemma \ref{lemma:main-lemma}, as \proName\ satisfies all three conditions (\textsf{C1-C3}) of the Lemma.
\end{proof}
\begin{lemma}
\label{lemma:main-lemma}
Let $A_{fr}$ be the class of semi-honest, front-running adversaries. For each adversarial strategy $A_{2}$, these exists adversarial strategy $A_{1} \in A_{fr}$,  $$U(\Pi,A_{1}) + negl(\kappa) \geq U(\Pi,A_{2})$$ and it is true when the following are satisfied: 
\begin{itemize}
    \item[{\small\textsf{C1}}] Reward-scheme and externality is such that it satisfies \emph{All-honest-profitability}.
    \item[{\small\textsf{C2}}] $\beta_{adv} < \frac{1}{2}$
    \item[{\small\textsf{C3}}] The block-reward scheme is inflationary.
\end{itemize}
\end{lemma}
\begin{proof}
This proof proceeds in 3 steps (7 sub-steps). In Step 1, we find the utility of a front-running adversary $A_{1} \in A_{fr}$. More specifically, we find the environment under which this adversary exists and the Reward $\mathcal{R}_{A_{1}}$ in Step 1a. Then in Step 1b, we find an appropriate lower bound on the probability of mining by $A_{1}$, after considering the variable difficulty and dynamic population. Finally, in Step 1c, we take into account variable block reward (inflationary) and find a lower bound on expected reward for $A_{1}$, or $E[\mathcal{R}_{A_{1}}]$.

In Step 2 of the proof, we upper bound the payoff of any other arbitrary adversary $A_{2}$ for its maximizing environment $Z_{2}$. In this case, we find the upper bound on the expected payoff of the adversary $A_{2}$. Then in Step 3a, we argue that an environment $Z_{1}$ always exists for every $Z_{2}$, such that a condition holds true. We argue that under such an environment, except with negligible probability, the payoff of $A_{1}$ exceeds the expected payoff of $A_{2}$. Finally, in Step 3b, we argue by contradiction that the adversarial setting $(S_{1},Z_{1})$ is strongly attack-payoff secure. Where, $S_{1}$ is the ideal world simulator of $A_{1} \in A_{fr}$. (ref.Appendix~\ref{lemma:main-lemma-appendix} for complete proof)
\end{proof}
The \emph{all-honest-profitability} condition is to ensure non-deviating parties participate in the system. Further, \ap\ has incentives both internal (through coins), and external (through short position on the currency) and is therefore incentivized to participate irrespective of \emph{all-honest-profitability} condition. Since we proved this theorem for general diverging series $\vartheta$, this is true for series such as constant-series ($\vartheta(i) = c$), harmonic series ($\vartheta(i) = \frac{1}{i}$) etc.

\subsubsection{Results for Deflationary Currency}
\label{sssec:result-deflation}
One of the advantages of \proName\ is that even under a deflationary reward scheme, it provides strong attack payoff security against a class of adversaries whose attacks are bounded by the number of rounds. In this section, we first show that PoW blockchains with geometrically decreasing block-reward schemes (like Bitcoin) are not strongly attack-payoff secure against a PPT ITM adversary. We then show that such a PoW blockchain when following \proName\ framework, is strongly attack-payoff secure against a PPT ITM adversary with an upper limit on the number of rounds on their attack.

\begin{theorem}[Deflationary Reward Scheme]\label{thm:dra}
PoW blockchain with geometrically decreasing Deflationary Reward Scheme,  ($\vartheta(i) = \vartheta^{i} $ for $\vartheta < 1$) cannot be strongly attack-payoff secure in Block-Reward model. We assume the protocol initially (at $t=0$) satisfies \emph{all-honest-majority}.
\end{theorem}
\begin{proof}
This proof follows in three steps. In Step 1, we argue why the result is true when rewards do not satisfy all-honest-profitability (Definition~\ref{def:ahp}). For all-honest-profitability, the proof is further divided in Steps 2,3. In Step 2, we find an environment $Z_{2}$ for any adversary $A_{2}$ with a slight advantage in the probability of mining (such as due to selfish mining). In Step 3, we complete the proof by showing $A_{2}$ has a higher expected payoff in environment $Z_2$ than any front-running semi-honest adversary $A_{1}\in A_{fr}$. 
The complete proof is provided in Appendix~\ref{sssec:semi-major-proof}.\end{proof}
\begin{theorem}\label{thm:semi-major-theorem}
For attacks $A_{2}$ which extend for less than $\alpha_{th}$ phases, \proName\ with  deflationary ($\vartheta$ is geometrically decreasing) block-reward scheme is strongly attack-payoff secure against a computationally bounded adversary $A\in\mathcal{A}^{\alpha_{th}}$ for $\beta_{adv} < \frac{1}{2}$ where, 
$$\alpha_{th} = 1 + \lfloor \frac{log(1 - p_{fr})}{log(\vartheta)} \rfloor$$
Here $p_{fr}$ is the probability that the protocol accepts a query by a front-running semi-honest adversary.
\end{theorem}
\begin{proof}
Proof of this theorem uses the adversary discussed in Theorem~\ref{thm:dra}. This adversary is the smallest powerful adversary that can achieve a greater payoff from any front-running strategy. We bound the attacker to be weaker than this adversary to obtain the result. The complete proof is given in Appendix~\ref{app:semi-major-proof}
\end{proof}


\section{Conclusion and Future Work}
\noindent \textbf{Conclusion.} In this paper, we analyzed and found security attacks possible on blockchain protocols. E.g., Bitcoin is not secure against adversary control $.45$ fraction of the computing power. We identified reasons why previous security analysis models fail to capture these. Towards this, we proposed a novel model of Rational Protocol Design, \prpd. Using this, we designed solutions to address these attacks and proposed a framework for designing PoW blockchain protocols, namely, \proName. We proved that \proName\ is strongly attack-payoff secure under an inflationary block-reward scheme. Under a deflationary block-reward scheme, we prove that \proName\ is secure against an adversary bounded by the number of rounds. \\

\noindent\textbf{Future Work.} The model used for security analysis of PoW blockchain protocol fails to capture rational deviations which are incentivized from outside the system, such as the attacks proposed in~\cite{Judmayer-2021-AIM}. We believe such attacks can be captured through the generalization of \prpd. We believe our results expand the existing models of Game-Theoretic security of Blockchains to a more general model. Extension of models of security for other types of blockchain protocols, such as PoS and other cryptographic protocols against incentive-driven adversaries, might be of interest and is left for future work.

\newpage

\bibliographystyle{unsrtnat}

\newpage


\appendix
\section{Proofs of Theorems Regarding Attacks}

\subsection{Proof of Theorem~\ref{thm:daa-attack}}
\label{ssec:app-daa-attack}
\underline{Step 1}:
Note that the utility of the adversary is
$$U_{A} = \max_{\mathcal{Z}\in ITM}\min_{\mathcal{S}\in C_{A}} E[v_{A}^{\mathcal{G}_{weak-ledger},\mathcal{S},\mathcal{Z}}]$$
Since the objective of \pd is to ensure security in the worst-case, we consider an environment that maximizes $U_A$. The adversary optimally chooses the parameters ($\alpha,r_1,r_2)$ under its control as follows:
\begin{itemize}
    \item[\ding{202}] to maximize the probability of successful attack, the adversary maximizes the duration, i.e., it sets $r_1=0,r_2=1$. 
    \item[\ding{203}] For the adversary, to launch attack, it is a best strategy is to adjust $\alpha$ such that $\tau_{adv} = \tau_{min}$ which is achieved when it sets $\alpha=\frac{\beta_{adv}}{\tau_{min}}$.
\end{itemize}
With these parameters, Equation~\ref{eq:daa-time} reduces to\\
$\frac{\tau_{hon}}{1 - \beta_{adv}} + \frac{1}{1 - \beta_{adv}} > \frac{1}{\beta_{adv}} + \frac{\tau_{adv}}{\beta_{adv}}$
$\Rightarrow \frac{\beta_{adv}}{1 - \beta_{adv}} > \frac{1 + \tau_{min}}{1 + \tau_{hon}}$\\
\underline{Step 2}: 
Our analysis only concerns $\beta_{adv} < \frac{1}{2}$ because for $\beta_{adv} > \frac{1}{2}$ forking using the $51\%$ attack~\cite{51P} is always possible. Additionaly, the attack being a security attack, \eo\ observes the attack and thus, by definition \rp\ follows honest strategy and we treat them as honest. We therefore show the result for $\tau_{min} < \frac{1}{2}\Rightarrow 1 - \beta_{adv} > \frac{1}{2} > \tau_{min}$.  \\
\underline{Step 3}: With the parameters set as described in Steps 1 and 2, we have the following inequality: 
$$\frac{2\beta_{adv} - \beta_{adv}^{2} - 1 + \beta_{adv}}{1 - \beta_{adv}} > \tau_{min}$$
On solving for $\beta_{adv}$, we have
$$\beta_{adv}^{2} - (3 + \tau_{min})\beta_{adv} + (1 + \tau_{min}) < 0$$
The roots of the equation are $L_{1}, L_{2}$, where 
$$L_{1} = \frac{(3 + \tau_{min}) - \sqrt{(3 + \tau_{min})^{2} - 4(\tau_{min} + 1)}}{2}$$
$$L_{2} = \frac{(3 + \tau_{min}) + \sqrt{(3 + \tau_{min})^{2} - 4(\tau_{min} + 1)}}{2}$$
The feasible region for $\beta_{adv}$ is $(L_{1},L_{2})$. However, $L_{2} > 1$, so the intersection of possible $\beta_{adv}$ values with values feasible for \daa\ attack gives us the bound $\beta_{adv} \geq \frac{(3 + \tau_{min}) - \sqrt{(3 + \tau_{min})^{2} - 4(\tau_{min} + 1)}}{2}$.\\
\underline{Step 4}:
Let $Q_{i}^{adv}$ and $Q_{i}^{hon}$ be random variables (RV) denoting the time taken to mine one block by the adversary and the \hp\ respectively. We define two RVs $Q^{adv}$ and $Q^{hon}$ as follows. $Q^{adv} = \sum_{i=0}^{2*\lambda} Q^{adv}_{i}$ and $Q^{hon} = \sum_{i=0}^{2\lambda} Q^{hon}_{i}$. $Q^{adv}$ and $Q^{hon}$ denote the total time to mine $2*\lambda, \mbox{blocks;}$. The factor $2$ is because attack progresses for $2$ phases (since $1-r_1=r_2=1$). Since RVs $Q_{i}^{adv}$s (similarly $Q_i^{hon}$) are independent of each other for different values of $i$, we can apply Chernoff bound (from~\cite{Upfal-Probability} Equation~4.2 and Equation~4.5)
$$Pr[Q^{adv} \geq (1 + \varepsilon)E[Q^{adv}]] < e^{-\frac{E[Q^{adv}]\varepsilon^{2}}{3}}$$
$$Pr[Q^{adv} \leq (1 - \varepsilon)E[Q^{adv}]] < e^{-\frac{E[Q^{adv}]\varepsilon^{2}}{2}}$$
Summing up the deviation probabilities, the expected time to mine $\lambda(r_{1} + r_{2})$ blocks deviates by more than $\varepsilon$ with probability $negl(\Theta\varepsilon^{2})$, as Expected time to mine a block is $\propto \Theta$. Therefore, 
the \daa\ Attack is successful with probability $>1- 2\cdot negl(\Theta\varepsilon^{2})$.

\subsection{Proof of Thm.~\ref{thm:pcf-attack}}
\label{app:pcf-attack-proof}
We prove this in four steps by calculating payoffs on deviating and following the protocol (Step 1 \& 2). Then comparing them to derive the bound (Step 3) and finally calculating the probability of success of this attack (Step 4).

\noindent \underline{Step 1} \textbf{Payoff on deviating:} 
On deviation, the last $k$ blocks of $P_{i}$ of chain $C_{H}$ are orphaned, and the mining cost spent by the \ap, \rp\ and \hp\ on the main chain is wasted for these blocks. This reduces $k\beta_{par}\chi_{1}$ from the payoff for $par \in \{rat, adv\}$. In addition, Mining on the $C_{A}$ incurs cost = $\frac{(k+m)\beta_{par}\chi_{1}}{\beta_{rat} + \beta_{adv}}$. This is higher as \hp\ are not mining on $C_{A}$.  In addition, the block reward collected by $par$ is $\frac{\beta_{par}}{\beta_{rat} + \beta_{adv}}$ of the total block reward of these $k + m$ blocks in the main chain. This value is equal to $\frac{(k + m\vartheta)r_{block}\beta_{par}}{\beta_{adv} + \beta_{rat}}$. Combining all three, we get the expected payoff on deviating as: 
\begin{equation*}
\resizebox{0.95\linewidth}{!}{$v_{par}(\pi^{'}) = \frac{n-1}{n}\frac{\beta_{par}}{\beta_{rat} + \beta_{adv}}\big(r_{block}(k + m\vartheta) - \chi_{1}(k+m)\big) - k\beta_{par}\chi_{1}$}
\end{equation*}
\pcf\ Attack is successful w.p. $\frac{n-1}{n}$ which we show in Step 4 of the proof.

\noindent \underline{Step 2} \textbf{Payoff on following:} On following the protocol, the payoff for $par$ (\rp\ or \ap) is $r_{block}(k + m\vartheta)\beta_{par}$ and the cost incurred is $\chi_{1}(k + m)\beta_{par}$. Therefore, the total payoff is:
$$v_{par}(\pi) = \beta_{par}\big(r_{block}(k + m\vartheta) - (k+m)\chi_{1}\big)$$
We can rewrite $v_{par}(\pi^{'})$ as $v_{par}(\pi^{'}) = \frac{n-1}{n}(\frac{v_{par}(\pi)}{\beta_{rat} + \beta_{adv}} - \chi_{1}k\beta_{par})$. Further, we substitute $\eta = \frac{r_{block}}{\chi_{1}}$

\noindent \underline{Step 3}: The condition $v_{par}(\pi^{'}) - v_{par}(\pi) > 0$ 
\begin{equation*}
\Rightarrow\chi_{1}\beta_{par}(\frac{(n\beta_{hon}-1)}{n(1 - \beta_{hon})})((\eta - 1)k + (\eta\vartheta - 1)m > \chi_{1}\beta_{par}k
\end{equation*}
\begin{equation*}
\Rightarrow (\frac{(n\beta_{hon}-1)}{n(1 - \beta_{hon})})((\eta - 1)k + (\eta\vartheta - 1)m > k
\end{equation*}
Since $C_{A}$ overtakes $C_{H}$ by $m$ blocks of phase $P_{i+1}$, therefore, the time taken to mine $m$ blocks in $C_{H}$ by $\beta_{hon}$ mining power is $\geq$ as the time taken to mine $k + m$ blocks in $C_{A}$ by $\beta_{rat} + \beta_{adv}$ mining power. For notational ease, we represent $J = \frac{\beta_{hon}}{1 - 2\beta_{hon}}$.  Thus,  $$\frac{m}{\beta_{hon}} \geq \frac{m + k}{1 - \beta_{hon}} 
\Rightarrow m \geq \frac{k\beta_{hon}}{1 - 2\beta_{hon}} = kJ$$
Clearly, as $m > 0$, $\beta_{hon} < \frac{1}{2}$,implying $\beta_{adv} + \beta_{rat} > \frac{1}{2}$. We take the earliest possible $m$, which gives us $m = \frac{k\beta_{hon}}{1 - 2\beta_{hon}}$. This gives us $v_{par}(\pi^{'}) - v_{par}(\pi) > 0$ as 
$$\Rightarrow (\frac{(n\beta_{hon}-1)}{n(1 - \beta_{hon})})(\eta - 1+ J\eta\vartheta - J) > 1$$
after substituting $J$ and simplifying, we get
$$ \eta > \frac{1 - \beta_{hon}}{n\beta_{hon} - 1}\cdot\frac{n - \beta_{hon}n - 1}{1 - (2 - \vartheta)\beta_{hon}}$$
\noindent \underline{Step 4}: 
For the attack to be successful, the lead of \hp\ should drop from $k$ to $0$ before it reaches $\rho$. This can be solved as Gambler's ruin problem (Sec 4.5.1~\cite{Ross-Probability}) with random walk moving in favor of \hp\ with probability $\beta_{hon}$. With this, the probability of an attack is 
$\frac{1 - \phi^{\rho - k}}{1 - \phi^{\rho}}$
 where $\phi = \frac{\beta_{hon}}{\beta_{rat}+\beta_{adv}}$. This probability is greater than $\frac{n-1}{n}$ if.
$
\frac{1 + (n-1)\phi^{\rho}}{n} > \phi^{\rho - k}$. We can simplify this as
\begin{equation}
    k < \lfloor \rho - log_{\phi}(\frac{1 + (n-1)\phi^{\rho}}{n}) \rfloor \label{eqn:k-lim}
\end{equation} 
With this, $k$ the probability of the \pcf\ attack being successful is at least $> \frac{n-1}{n}$.

\subsection{Proof of Thm.~\ref{thm:smb-attack}}
\label{app:smb-proof}
\begin{proof}
For proof of this attack, we consider that the bribe amount is a $z$ fraction of the Block reward for a single block ($z>0$, but a small value). The behavior of \ap, \rp and \hp\ follows as described in the Attack-strategy in Section~\ref{sssec:smb-attack}. In case of a tie between \ap\ and \hp\ blocks, if \ap's mined block becomes part of the longest chain, the payoff is $(1 - z)r_{block}$ for the adversary, and the party which mines block on top of \ap\ block gets payoff $(1 + z)r_{block}$. 

Bribes incentivize \rp s to deviate from the protocol. 
Due to this, it is safe to consider the same payoffs as in~\cite{Eyal-2014-Selfish}; however,  
the $\gamma$ -- the fraction of non-adversarial parties mining on the adversarial block in the case of a tie for the longest chain is $\frac{\frac{\beta_{h}}{2} + \beta_{r}}{\beta_{h} + \beta_{r}}$. Using this value in the result from~\cite{Eyal-2014-Selfish}, also given in Equation~\ref{eqn:selfish-mining-eyal}, we get 
$\frac{1}{2} > \beta_{a} >\frac{\beta_{h}}{2\beta_{r} + 4\beta_{h}}$\end{proof}

\subsection{Proof for Lemma~\ref{lemma:gossip-subopt}}
\label{app:gossip-subopt-proof}
\begin{proof}
First, we calculate the utility for a rational party $i$ for following the protocol $\Pi_{Gossip}$, i.e., broadcasting a transaction $\mathfrak{T}$ with transaction fee $tx_i$ that it hears. Then, we calculate its utility for not broadcasting $\mathfrak{T}$. We account for the unfavourable events (unfavourable for the attacker) that (i) some other parties may add $\mathfrak{T}$ and (ii) discounting the rewards if $\mathfrak{T}$ is added by the party later. We then argue that later leads to a higher utility. 
\\
\noindent\underline{Broadcasting $\mathfrak{T}$}
The probability of a single party mining a block in round $t$ is $p_{su}(t)$, and they can make $q$ queries in each round. Then, the utility in following the gossip protocol is given for a party with $\beta_{i}$ fraction of mining power in their control as : 
$$u_{i}^{*}(\beta_{i}) = (\sum_{j=1}^{n} tx_{j})(\sum_{t=1}^{\infty}\delta^{t} p_{su}(t)(1-p_{su}(t))^{n(t)\cdot t})$$
\noindent{\underline{Not Broadcasting $\mathfrak{T}$}} The utility for party $i$ becomes, 
$$u_{i}^{'}(\beta_{i}) = (\sum_{j=1,j\neq i}^{n} tx_{j})(\sum_{t=1}^{\infty}\delta^{t} p_{su}(t)(1-p_{su}(t))^{n(t)\cdot t}) + \frac{tx_{i}}{p_{su}}$$
As $\mathfrak{T}$ is accepted with probability $p_{su}$ in each round. Consider $K$ is the random variable $1$ when the party mines a block and $0$ otherwise. Then, $K$ is a geometric random variable with $E[K] = \frac{1}{p_{su}}$. We can clearly see that $u^{*}_{i}(\beta_{i}) < u^{'}_{i}(\beta_{i})$. Therefore $\sigma_{tw}$ dominates $\Pi_{gossip}$. 
\end{proof}

\section{Proofs of Detering Attacks}
\subsection{Proof for Lemma~\ref{lemma:smb-impossibility-lemma}}
\label{app:smb-impossibility-lemma}
\begin{proof}
Our proof proceeds in three steps. In Step 1, we define an environment $(Z_{1},S_{1})$ for $S_{1} \in \mathcal{A}_{fr}$. In Step 2, we define $(Z_{2},S_{2})$ where $S_{2} \in \mathcal{A}_{SMB}$. We then show in Step 3 that both these executions are indistinguishable from each other. \\

\noindent \underline{Step 1}: Consider a simulator $S_1 \in \mathcal{A}_{fr}$ be a semi-honest adversary. Environment $Z_{1}$ is such that it observes all chains, and if there is a contest between two chains that are at the same height, they maximally delay messages from miners mining on the chain with the last block not mined by party corrupted by $S_1$. 

\noindent \underline{Step 2}: Consider any simulator simulating \smb\ attack ($S_{2} \in \mathcal{A}_{SMB}$) and any general environment $Z_{2}$, which communicates messages in the same (partially-synchronous) manner for both \ap, \rp\ and \hp. 

\noindent \underline{Step 3:} It is clear by comparison that for any party viewing the two systems, $E[v^{\Pi,S_{1},Z_{1}}] \equiv E[v^{\Pi,S_{2},Z_{2}}]$, where $v$ represents the external view of the system. This means that for any PPT ITM it is not possible to distinguish between the two systems. Let $D$ be the discriminator which is a PPT ITM. Let us denote two systems $\rho_{1} = <\Pi,S_{1},Z_{1}>$ and $\rho_{2} = <\Pi,S_{2},Z_{2}>$. We denote a random variable $C$ which can take value $1$ and $2$ with equal probability. We select a system $\rho = \rho_{C}$ to be sent to the discriminator $D$ based on the value of $C$. Then, if $D$ is a PPT ITM, then $Pr[D(\rho) = C] = \frac{1}{2} + negl(\varepsilon)$.\\

Thus, because the two environments are indistinguishable we cannot ensure attack-payoff security (adversary follows $A \in \mathcal{A}_{fr}$ without also allowing \smb. 

\begin{figure}
\centering
\begin{tcolorbox}
\begin{center}
    $\mathcal{Z}^{Network}_{1}$
\end{center}
\large{on \textsc{RECEIVE()}}\

\noindent \hspace{0.3cm}$M\gets \{0,1\}^{*}$ (receive broadcasted message)

\noindent \hspace{0.3cm} if sender($M$) is $\mathcal{A}$

\noindent \hspace{0.6cm}- $B_{\mathcal{A}}\gets B_{\mathcal{A}}\cup\;M$

\noindent \hspace{0.3cm} else

\noindent \hspace{0.6cm} - broadcast first $max(2,|B_{\mathcal{A}}|)$ messages of $B_{\mathcal{A}}$

\noindent \hspace{0.6cm} - broadcast $M$
\end{tcolorbox}
\caption{Environment $Z_{1}$}
\label{fig:environment-Z1}
\end{figure}








\end{proof}

\subsection{Proof of Lemma~\ref{lemma:gossip-sol}}
\label{ssec:app-txincl-proof}

Let $Pool$ be the group of parties collectively deviating from \txincl\ and withhold a set of transactions with a cumulative transaction fee $tx$. Let $\beta_{Pool}$ be their collective computing power. 

First, we focus on  and compare the utility $Pool$ obtains from following gossip protocol versus \tw. Next, since the analysis is on the transaction network, we assume the PoW protocol to be followed correctly. For the sake of abstraction, we consider the probability of mining a block in a round $t$ by a single party is $p_{su}(t)$ and the total number of parties are $n(t)$. The payoff on following the gossip protocol becomes : 
$$v(\Pi_{gossip}) = \sum_{i=1}^{\infty} \delta^{i-1}v_{i}(\Pi_{gossip})$$
where $\delta$ is the discount factor, which captures the increasing chance of the proposer sending the transaction to another party, thereby reducing the chance of the current party exclusively mining for that transaction. $v_{i}$ is the expected payoff from the transaction in the $i^{th}$ round.

$v_{i}(\Pi_{gossip}) = $
\begin{equation*}
(1 - (1 - p_{su}(i))^{\beta_{Pool}n(i)})(1 - p_{su}(i))^{n(i)\cdot(i-1)}\frac{tx}{2^{l}}
\end{equation*}
On summing up $\sum_{i=0}^{\infty}v_{i}(\pi_{gossip})$, we get
\begin{equation*}
\sum_{i=0}^{\infty} \delta^{i}(1 - (1 - p_{su}(i+1))^{\beta_{Pool}n})(1 - p_{su}(i+1))^{n(i+1)\cdot i}\frac{tx}{2^{l}}
\end{equation*}
For the deviating protocol ($\Pi_{secret}$), the utility in the $i^{th}$ round $v_{i}(\Pi_{secret})$ becomes, 
$$(1 - (1 - p_{su}(i))^{\beta_{Pool}n(i)})(1 - p_{su}(i))^{n(i)\cdot(i-1)\beta_{Pool}}\frac{tx}{2^{l}}$$
We therefore have 
$$v_{i}(\Pi_{secret}) - v_{i}(\Pi_{gossip}) = (1 - (1 - p_{su}(i))^{\beta_{Pool}n(i)})\frac{tx}{2^{l}}$$
$$(1 - p_{su}(t))^{n(i)\cdot(i-1)\beta_{Pool}}(1 - (1 - p_{su}(t))^{n(i)\cdot(i-1)(1 - \beta_{Pool})})$$
If we sum this over geometrically decreasing $\delta$, we get 
$$v(\Pi_{secret}) - v(\Pi_{gossip}) = \sum_{i=1}^{\infty}(1 - (1 - p_{su}(i))^{\beta_{Pool}n(i)})\frac{tx}{2^{l}}\mathfrak{R}$$
$$\mathfrak{R} = (1 - p_{su}(i))^{n(i)\cdot(i-1)\beta_{Pool}}(1 - (1 - p_{su}(i))^{n(i)\cdot i(1 - \beta_{Pool})})$$
Taking upper limit of probability as $1$ for $(1 - (1 - p_{su}(i))^{n(i)\cdot i(1 - \beta_{Pool})})$, we get 
$$\mathfrak{R} \leq (1 - p_{su}(i))^{n(i)\cdot(i-1)\beta_{Pool}} = \mathfrak{M}$$
Therefore $\mathfrak{M}$ is the probability that no miner mines a block at round $i$. 
$$v(\Pi_{secret}) - v(\Pi_{gossip}) \leq \sum_{i=1}^{\infty}(1 - (1 - p_{su}(i))^{\beta_{Pool}n(i)})\frac{tx}{2^{l}}\mathfrak{M}$$
If we upper bound the remaining probability term to $1$, we get the expression 
$$ v(\Pi_{secret}) - v(\Pi_{gossip}) \leq \frac{tx}{2^{l}}\sum_{i=0}^{\infty} \mathfrak{M}$$
Since total probability of the block not being mined ($\sum_{i=1}^{\infty} \mathfrak{M} \leq 1$) we get the bound 
$$ v(\Pi_{secret}) - v(\Pi_{gossip}) \leq \frac{tx}{2^{l}} $$
    

Since $tx$ is polynomial in $l$, $\frac{tx}{2^{l}}$ is negligibly small. Let $\epsilon_{G} = O(tx\cdot2^{-l})$. We therefore get $v(\Pi_{secret}) - v(\Pi_{gossip}) \leq \epsilon_{G}$

\section{Proofs for PRAGTHOS Analysis}
The proofs for \ref{lemma:main-lemma} and \ref{thm:dra} are in the Attack model $<\mathcal{F},<\mathcal{F}>,\overline{v}>$ where as decribed in \ref{ssec:attack-model}, the attack model is $<\mathcal{G}_{ledger}, \mathcal{G}_{weak-ledger},$\\$ (v_{A},v_{R},v_{H})>$

\subsection{Proof for Lemma~\ref{lemma:main-lemma}}
\label{lemma:main-lemma-appendix}
\begin{proof}
\noindent\underline{Step 1a} Let $A_{1} \in A_{fr}$ be a front-running adversary which makes $q^{*}$ queries. Further consider the environment $Z_{1}$ which runs the execution of the protocol where the adversary is activated to make $q^{*}$ queries before halting, and the \hp s are activated till at least one of them output all the blocks mined by the adversary in their longest chain. We consider real-world UC execution, and all random variables are correspondingly defined. 

Consider random variable $X_{i}$ which is $1$ if $i^{th}$ query by adversary successfully mines a block, and $0$ otherwise. Thus, the payoff for \ap in $q^{*}$ queries is $$\mathcal{R}_{A_{1}} = \sum_{i=1}^{q^{*}} X_{i}r_{block}\theta - \chi$$ Notice that we exclude the payoff that the adversary gets by decreasing the value of coin by lowering $\theta(t)$ through security attacks -- forking the chain, because in \proName, forking is not possible for $\beta_{adv} < \frac{1}{2}$. 

\noindent\underline{Step 1b} Since due to variation in the number of miners, the difficulty of mining, and hence the probability of a block getting accepted changes with each epoch, the probability of getting a block accepted is $p_{e}$ in epoch $e$. As the number of miners do not grow exponential across rounds, we can assume that there exists a polynomial $s(t)$ , such that $n(t) \leq s(t)\;\forall t$. Let $p^{min}_{e}$ be the probability of $A_{1}$ mining a block in round $t$ if $n(t)$ grows exactly as $s(t)$ .  Clearly, $\forall$ epochs, $p_{e} \geq p^{min}_{e}$. Let $p_{min} = \min_{\forall e} p^{min}_{e}$.

\noindent\underline{Step 1c} Consider that the sum of the block-reward  up to $a$ queries is denoted by $\mathfrak{J}(a)\cdot r_{block}(0)$. $$\mathfrak{J}(a) := \sum_{i=0}^{a} \frac{r_{block}(i)}{r_{block}(0)} \approx \sum_{j=0}^{\alpha_{a}} \Lambda\vartheta(j)$$
We have replaced queries with number of blocks mined, because the expected number of blocks mined deviates very less for large number of queries, and in $q$ queries, the number of blocks mined does not deviate by more than negligible amount with overwhelming probability. This result follows from the Chernoff-bound analysis.
Since both $\theta$ and $\chi$ are variable for rounds $1$ to $r_{1}$ till which the protocol runs, we define $\eta_{max} = \max_{t \in [1,r_{1}]} \frac{\chi}{p_{e}\theta(t)r_{block}}$. Since mining is profitable, we can conclude that $\eta_{max} < 1$. Also $\theta = \theta(t)$ is same $\forall\;t$ because both $\Pi$ and Adversarial strategy is fixed. Now, we can conclude that $$E[\mathcal{R}_{A_{1}}] \geq \mathfrak{J}(q^{*})(1 - \eta_{max})p_{min}r_{block}\theta$$

\noindent\underline{Step 2}
Now, consider any arbitrary adversary $A_{2} \in ITM$. This adversary makes $Q$ queries during its execution in an environment $Z$. Let $P_{Q}$ be the distribution of the number of queries made by this adversary $q = \max \text{support}(P_{Q})$. Let $Z_{2}$ be environment where $Q = q$. Consider this $(A_{2},Z_{2})$, where the expected payoff of the adversary is upper bounded by taking probability of mining per query as $1$, which gives us $$E[\mathcal{R}_{A_{2}}] \leq \mathfrak{J}(q)r_{block}\theta$$
This upper bound comes from the fact that a single query can extend the blockchain by at most one block, in the $\mathcal{F}$ functionality. However, if that is not the case with it's UC-Realization, the state exchange protocol $\Pi$, then $EXEC_{\mathcal{F},\mathcal{S}_{2},\mathcal{Z}_{2}} \not \approx EXEC_{\Pi_{B},\mathcal{A}_{2},\mathcal{Z}_{2}}$, and thus $C_{A_{2}} = \phi$, which is not possible.

\noindent\underline{Step 3a} We choose $Z_{1}$ such that the condition $\mathfrak{J}(q^{*}) \geq \frac{\mathfrak{J}(q)\kappa}{(1 - \delta)(1 - \eta_{max})p_{min}}$ is satisfied. Because the crypto-currency is inflationary in nature, we are assured that such an environment always exists. This is because for inflationary series $\mathfrak{J}$, there always exist $a_{2} > a_{1}$ such that $\mathfrak{J}(a_{2}) > \mathfrak{J}(a_{1})$, $$\because\lim_{a_{2}\rightarrow \infty} \mathfrak{J}(a_{2}) - \mathfrak{J}(a_{1}) \longrightarrow \infty \; \forall \; a_{1}$$

Now, consider the probability that the payoff of $A_{1}$ is less than the expected payoff of $A_{2}$. We can say there always exist such $Z_{1}$ for each $S_{2}$, because there always exist such $q^{*}$ for $q$, due to the diverging nature of the series $\vartheta$.
$$Pr[\mathcal{R}_{A_{1}} < E[\mathcal{R}_{A_{2}}]] \leq Pr[\mathcal{R}_{A_{1}} < \mathfrak{J}(q)r_{block}\theta]$$
$$\leq Pr[\mathcal{R}_{A_{1}} < (1 - \delta)(1 - \eta_{max})\mathfrak{J}(q^{*})r_{block}p_{min}\theta]$$
$$\leq Pr[\mathcal{R}_{A_{1}} < (1 - \delta)E[\mathcal{R}_{A_{1}}]] = Pr[\sum_{i=1}^{q^{*}} X_{i} \leq (1 - \delta)E[\sum_{i=1}^{q} X_{i}]$$
$$< e^{-\frac{\delta^{2}q^{*}p}{2}}$$

\noindent The last inequality follows from Chernoff's bound~\cite{Upfal-Probability}. 
The difference in expected values of all Random variables \textbf{$(Q,\mathcal{R}_{A_1},\mathcal{R}_{A_2},X_{i}...)$} in both Real and Ideal execution cannot be   more then a small amount $\epsilon$. This is because, if the expected value deviates by more than $\epsilon$, this event is observable, and by Chernoff-bound analysis\footnote{$Pr[X_{ideal} \geq (1 - \epsilon)E[X_{real}]] < exp(-\frac{\epsilon^{2}E[X_{real}]}{2})$} it's value is very small. Therefore, such an event happening means that $EXEC_{\Pi_{B},A,Z} \not \approx EXEC_{\mathcal{F},S,Z}$ with very high probability; which contradicts that $S \in C_{A}$. Therefore, for ideal payoff $v_{A}$, we have $E[v_{A}^{\mathcal{F},S_{1},Z_{1}}] \approx E[\mathcal{R}_{A_{1}}]$.

Using this, we can conclude the result that $E[v_{A}^{\mathcal{F},S_{1},Z_{1}}] \geq S(q)r_{block}\theta \geq E[v_{A}^{\mathcal{F},S_{2},Z_{2}}]$ with overwhelming probability.

\noindent\underline{Step 3b} Now, we need to show $U_{A}(\Pi,\mathcal{A}_{1}) + negl(\kappa) \geq U_{A}(\Pi,\mathcal{A}_{2})$. Let us assume this is false. That means $\exists S_{2},Z_{2}$ such that
\begin{center}
    $\max_{Z_{2} \in ITM}\Big\{\min_{S_{2} \in C_{A_{2}}}\Big\{ E[v_{A}^{\mathcal{F},S_{2},Z_{2}}]\Big\}\Big\} \geq$\\
    $\max_{Z_{1}\in ITM}\Big\{\min_{S_{1} \in C_{A_{1}}}\Big\{ E[v_{A}^{\mathcal{F},S_{1},Z_{1}}]\Big\}\Big\}$
\end{center}
This means that there exists a $Z_{2}$ such that $\forall Z_{1}$
$$\min_{S_{2} \in C_{A_{2}}} E[v_{A}^{\mathcal{F},S_{2},Z_{2}}] \geq \min_{S_{1} \in C_{A_{1}}} E[v_{A}^{\mathcal{F},S_{1},Z_{1}}]$$
But we have shown that this is false, except with negligible probability. This means that our assumption was wrong, and in reality, $\forall A_{2} \exists A_{1}$ such that the attack-payoff security condition holds.\end{proof}

\subsection{Proof for Theorem~\ref{thm:dra}}
\label{sssec:semi-major-proof}
\begin{proof}
It should be noted that we discuss this scheme for geometrically decreasing converging series, because this is the series that is employed in most of the existing PoW blockchains. But, the result holds for any converging $<\vartheta>$. 

\noindent \underline{Step 1 (No all-honest-profitability)}: Since the protocol initially (when started) follows all honest profitability, for this case, the condition $r_{block}(t)p_{hon}(t)\theta(t) < \chi$ is becomes true after some rounds. The dominant strategy for \hp\ and \rp\ is to abstain from the protocol. However, for \ap, this scenario could still be profitable by shorting the cryptocurrency (following $\mathcal{A}_{\overline{ahp}}$). 

\begin{tcolorbox}
\begin{center}
    $\mathcal{A}_{\overline{ahp}}$
\end{center}
\begin{enumerate}
    \item Hold \emph{short position} against the cryptocurrency in some round when ahp\footnote{all-honest-profitability} is satisfied.
    \item When $\overline{ahp}$, $\beta_{adv} = 1$, because all \rp\ and \hp\ leave the protocol. 
    \item When $\beta_{adv} > \frac{1}{2}$, launch security attack against the protocol bringing the value of the crypto-currency down and profit from the short position held. 
\end{enumerate}
\end{tcolorbox}

\noindent \underline{Step 2 (All-honest-profitability)}: For protocol to be not attack-pay-off-secure, we need to show the existence of an attack strategy $A_{2}$, which is simulated by simulator $S_{2}$ in environment $Z_{2}$, such that $\nexists A_{1} \in A_{fr}$ for any environment $Z_{1}$ for which the following equation holds $$E[v_{A}^{\mathcal{F},S_{1},Z_{1}}] < E[v_{A}^{\mathcal{F},S_{2},Z_{2}}]$$

Here, all, $A_{1}, A_{2}, S_{1}, S_{2}, Z_{1} \& Z_{2}$ are PPT ITMs\footnote{probabilistic polynomial time interactive Turing machines} .Let us consider any attack where the probability of acceptance of the block by the adversary per query is $p_{attack}$, which is greater than the same probability for a front-running semi-honest miner $p_{fr}$. Note that at least \smb, $p_{attack}>p_{fr}$. Our proof being general, we consider any attack where this is true. Now, for the attack-strategy $A_{2}$, which is simulated by simulator $S_{2} \in C_{A_{2}}$, and makes $q$ queries, which are distributed across $\alpha_{opt}$ phases. Consider the expected payoff for this adversary 
$$E[\mathcal{R}_{A_{2}}] = \Lambda\theta r_{block}p_{attack}\sum_{i=0}^{\alpha_{2}-1} \vartheta^{i} - q\chi $$
$$= \Lambda\theta r_{block}p_{attack}\frac{1 - \vartheta^{\alpha_{2}}}{1 - \vartheta} - q\chi$$
Now consider any front-running adversary $A_{1} \in A_{fr}$, which is simulated by $S_{1} \in C_{A_{1}}$ and the environment $Z_{1}$ which allows the adversary to make $q^{*}$ queries and waits for at least \hp\ to output the chain with all adversarial blocks in it, before halting the execution. Consider that in expectation $\Lambda\alpha_{1}$ blocks are mined in this duration. The expected payoff of such an adversary is 
$E[\mathcal{R}_{A_{1}}] = \Lambda\theta r_{block}p_{fr}\sum_{i=0}^{\alpha_{1}-1} \vartheta^{i} - q^{*}\chi$, i.e., 
$$E[\mathcal{R}_{A_{1}}] = \Lambda\theta r_{block}p_{fr}(\frac{1 - \vartheta^{\alpha_{2}}}{1 - \vartheta} + \vartheta^{\alpha_{2}}\sum_{j=0}^{\alpha_{1}\vartheta^{j} - \alpha_{2}}) - q^{*}\chi$$
Since $p_{attack} > p_{fr}$, let $p_{attack} - p_{fr} = \Delta$. We want $\alpha_{opt}$ as the optimal $\alpha_{2}$ for which the expected payoff for $A_{2}$ is higher than that for $A_{1}$ for all $A_{1},Z_{1}$. Since the result holds $\forall\; Z_{1}$, we consider an environment where the cost of mining is negligible because in this environment, the adversary $A_{1}$ can make an arbitrary number of queries without incurring additional cost and is best suited for $A_{1}$ because $q \leq q^{*}$. Further, since the condition should be true $\forall \; A_{1}$, we take $\lim_{\alpha_{1}\rightarrow\infty}$, using which we get the condition 
$$(p_{fr} + \Delta)\frac{1 - \vartheta^{\alpha_{2}}}{1 - \vartheta} > p_{fr}\frac{1 - \vartheta^{\alpha_{2}}}{1 - \vartheta} + \frac{p_{fr}\vartheta^{\alpha_{2}}}{1 - \vartheta}$$
Let $\alpha_{opt}$ be the minimum $\alpha_{2}$  that  satisfies the above condition 
$$\Delta\frac{1 - \vartheta^{\alpha_{2}}}{1 - \vartheta} \geq \frac{p_{fr}\vartheta^{\alpha_{2}}}{1 - \vartheta}\Rightarrow 1 - \vartheta^{\alpha_{2}} > \frac{p_{fr}}{\Delta}\vartheta^{\alpha_{2}}$$
$$\Rightarrow\vartheta^{\alpha_{2}} < \frac{\Delta}{p_{fr} + \Delta}\Rightarrow\alpha_{2}log(\vartheta) < log(\frac{\Delta}{p_{fr} + \Delta})$$
Since $log(\vartheta) < 0$, and we want to minimize $\alpha_{2}$, we can say 
$$\alpha_{opt} = 1 + \lfloor \frac{log(\frac{\Delta}{p_{fr} + \Delta})}{log(\vartheta)} \rfloor$$
$$E[\mathcal{R}_{A_{2}}] = \Lambda\theta r_{block}p_{attack}\frac{1 - \vartheta^{\alpha_{opt}}}{1 - \vartheta} - \Lambda\alpha_{opt}\chi$$
$$E[\mathcal{R}_{A_{1}}] = \Lambda\theta r_{block}p_{fr}\sum_{i=0}^{\alpha_{1}} \vartheta^{i} - q^{*}\chi$$
$$E[\mathcal{R}_{A_{1}}] = \Lambda\theta r_{block}p_{fr}(\sum_{i=0}^{\alpha_{opt}} \vartheta^{i} + (\vartheta^{\alpha_{opt}}\sum_{i=0}^{\alpha_{1} - \alpha_{opt}}\vartheta^{i})) - q^{*}\chi$$
\noindent \underline{Step 3}: Let us take simulator $S_{2}$ and environment $Z_{2}$ for which $\alpha_{2} = \alpha_{opt}$. Let $D=\mathcal{R}_{A_{2}} - \mathcal{R}_{A_{1}}$. We also lower bound the difference $(q^{*} - \Lambda\alpha_{opt})\chi = 0$. So$ E[D]$ is lower bounded by $$E[D] \geq \Lambda\theta r_{block}\Big(\frac{1 - \vartheta^{opt}}{1 - \vartheta}(p_{attack} - p_{fr}) + \frac{\vartheta^{opt}p_{fr}(1 - \vartheta^{d})}{1 - \vartheta}\Big)$$
Let $d = \alpha_{1} - \alpha_{opt}$ ($d \geq 0$) and $\Delta = p_{attack} - p_{fr}$. $$\Rightarrow E[D] \geq \Lambda\theta r_{block}\Big(\frac{1 - \vartheta^{\alpha_{opt}}}{1 - \vartheta}\Delta - \vartheta^{\alpha_{opt}}p_{fr}\frac{1 - \vartheta^{d}}{1 - \vartheta}\Big)\geq0$$
The last inequality follows from the definition of $\alpha_{opt}$.
Since we solved this for arbitrary $A_{1},Z_{1}$ (by showing result is true $\forall\; d \geq 0$), we conclude that $A_{2}$ is such that $S_{2}$, which simulates it in environment $Z_{2}$ has no $A_{1} \in A^{\infty}_{fr}$ such that $(S_{1},Z_{1})$ achieves higher payoff for any $Z_{1}$.

Further, we assumed that $\Delta > 0$, so even if there exists an attack that gives an adversary a slightly higher mining probability (e.g., \smb), the protocol is not strongly attack-payoff secure.
\end{proof}

\subsection{Proof for theorem \ref{thm:semi-major-theorem}}
\label{app:semi-major-proof}
\begin{proof}
In this theorem, we show that any deflationary block-reward based cryptocurrency is attack-payoff secure against the set of adversarial strategies bound to $\alpha_{opt}$ rounds. That is, $\forall\; A \in \mathcal{A}^{\alpha_{opt}}$
$$\alpha_{opt} = 1 + \lfloor \frac{log(1 - p_{fr})}{log(\vartheta)} \rfloor$$
Consider any adversary $A_{2} \in \mathcal{A}^{\alpha_{opt}}$ with environment $Z_{2}$ where it makes $Q = q$ queries, for $q = max support(P_{Q})$. $P_{Q},Q$ are as explained in proof for Lemma~\ref{lemma:main-lemma}. In this case, let $q$ queries mine blocks such that in expectation $\alpha_{2}$ phases are completed for $\alpha_{2} < \alpha_{opt}$. The payoff for the adversary $R_{A_{2}}$ is upper bounded by taking the probability of each query leading to a block being mined as $1$. 
$$R_{A_{2}} \leq \Lambda r_{block}\theta\sum_{i=0}^{\alpha_{2}-1}\vartheta^{i} = \Lambda\theta r_{block}\frac{1 - \vartheta^{\alpha_{2}}}{1 - \vartheta}$$ 
Consider a front-running semi-honest adversary $A_{1}$ and an environment $Z_{1}$ where the adversary makes $q^{*}$ queries before the environment halts. In this case, consider $p_{fr}$ be the probability of each query being accepted. Consider $q^{*}$ such that it runs for $\alpha_{1}$ rounds in expectation. The relation be such that $(1 - \vartheta^{\alpha_{1}}) \geq \frac{(1 - \vartheta^{\alpha_{opt}})\kappa}{(1 - \delta)(1 - \eta_{max})p_{min}}$, where all terms are same as defined in Step 1b in Appendix~\ref{lemma:main-lemma-appendix}. In this case, the adversary $A_{1}$ has payoff $$R_{A_{1}} \geq \Lambda\theta r_{block} \sum_{i=0}^{\alpha_{1}-1} \vartheta^{i} - \Lambda\theta r_{block}\frac{1 - \vartheta^{\alpha_{1}}}{1 - \vartheta}$$ 
$P[R_{A_{1}} < R_{A_{2}}] \leq P[\Lambda\theta r_{block}p_{fr} \frac{1 - \vartheta^{\alpha_{1}}}{1 - \vartheta} < \Lambda\theta r_{block}\frac{1 - \vartheta^{\alpha_{2}}}{1 - \vartheta}]$
$$= P[p_{fr}\frac{1 - \vartheta^{\alpha_{1}}}{1 - \vartheta} < \frac{1 - \vartheta^{\alpha_{2}}}{1 - \vartheta}] = P[\frac{p_{fr}}{1 - \vartheta^{\alpha_{2}}} < \frac{1}{1 - \vartheta^{\alpha_{1}}}]$$
There always exist such $\alpha_{1}$ for each $\alpha_{2} < \alpha_{opt}$, this probability is $1$ for some $Z_{1}$. (notice that we calculated $\alpha_{opt}$ in  in~\ref{sssec:semi-major-proof} to ensure this holds true).\end{proof}

\section{Other Details}
\label{app:misc-def}
\subsection{Reward Mechanisms in Blockchains}

Blockchain Reward mechanisms have been studied in prior works~\cite{b14, Badertscher-2018-BitcoinRPD}. These reward mechanisms can be broadly categorized into two categories (1) Block Reward Mechanism (BRM) and (2) Transaction Fee Only Mechanism (TFOM). 

\paragraph{Block Reward Mechanism (BRM).} In the block-reward mechanism, the incentive for mining is provided through a special`coinbase' transaction. The cryptocurrency is deflationary if the block reward reduces every finite number of blocks mined, such that the sum of the net block reward is constant. In Bitcoin, Block-Reward halves every $210000$ block mined. The payoff from transaction fees in BRM is very small compared to Block Reward and does not lead to strategic deviations. 

\paragraph{Transaction Fee Only Mechanism (TFOM).} In the TFOM, miners get negligible block rewards, and the main source of the payoff is the transaction fees from the transactions included in the blockchain. 

\subsection{Difference Utility}
\label{app:difference-utility}
In this section, we explain the advantage of Protocol descriptors having their utility modeled a difference in utility of deviating and non-deviating parties. \\

\noindent Modeling utility in this way allows us to capture cases where the adversary does not gain a significant increase in payoff, but it reduces the payoff of other parties per round. Doing so allows the adversary to capture more fraction of the block reward from a particular phase. Further, the representation of utility as the difference between payoffs of deviating and non-deviating parties is also a more practical representation of the goal of protocol descriptor which is to minimize the benefit that a party gets from deviating. 
\subsection{Goldfinger Attack}
\label{app:goldfinger}
Goldfinger attack~\cite{GoldFinger} is one of the attacks where the adversary holds a short-position of the cryptocurrency and then launches a security attack. In this case, they profit from the decrease in the conversion rate of the cryptocurrency. This can be modeled in the payoff of the adversary as 
$$v_{A} = \theta(t)E[R_{B}] + (\theta_{init} - \theta(t))c_{1}$$
Here $\theta_{init}$ is the coin's conversion rate when the short position was initially held. Therefore, as $\theta(t)$ decreases, the second term increases. This was not modeled in the utility structure of previous works such as~\cite{Badertscher-2018-BitcoinRPD}. Modeling this allows us to argue that even if mining is not profitable, the adversary can have a positive payoff by shorting the crypto-currency.  



\end{document}